\documentclass[]{aa}
\usepackage{graphicx}
\usepackage{lipsum}
\usepackage{txfonts}
\usepackage[colorlinks,citecolor=blue,urlcolor=blue,bookmarks=false,hypertexnames=true]{hyperref}

\begin{document}

   \title{Predictions for LISA and PTA based on SHARK galaxy simulations}
   \subtitle{}

   \author{M.Curyło
          \and
          T.Bulik
          }

   \institute{Astronomical Observatory, University of Warsaw, Al. Ujazdowskie 4, 00--478 Warsaw, Poland\\
              \email{mcurylo@astrouw.edu.pl}
            }

   \date{Received ...; accepted ....}

  \abstract{We present our analysis of a set of populations of massive black hole (MBH) binaries generated in the recent semi-analytic model of galaxy evolution (SHARK). We focus on studying gravitational wave (GW) emission produced during MBH mergers in terms of their detectability with current and future detectors, namely, Pulsar Timing Arrays  (PTAs) and  Laser Interferometer Space Antenna (LISA). The key advantage of SHARK is that it provides a way to explore a number of distinct models of MBH and galaxy evolution processes within a consistent framework and it was also successfully tested against current constraints from electromagnetic observations. In our work, we studied 12 models that vary in terms of their MBH seed formation scenarios and we tested two different MBH growth and feedback models.
  Based on our estimates, we find that LISA will be able to detect several to several tens of merger events per year for the most and least massive seed scenarios, respectively. We also show that the strength of this relation depends on the MBH growth model, where in the most extreme case, we find twice as many detected events for the same initial seed masses. Finally, we estimated the amplitude of the GW background at nHz frequencies to be on the order of $1.4\cdot10^{-16} - 1.1\cdot10^{-15}$. This value depends solely on the time delay between the merger of galaxies and their MBHs.
}


\maketitle


\section{Introduction}

Massive black holes (MBHs) residing in the centers of many (if not all) galaxies are fascinating objects which elude our basic
understanding even after decades of study. Indeed, numerous observational campaigns have contributed strong evidence to support their existence in the nuclei of both large galaxies \citep{Farrarese05} and also dwarf galaxies \citep{Reines13,Nguyen19,Baldassare20}, while also showing complex empirical relations between MBH masses and various properties of their host galaxies such as velocity dispersion, total stellar mass, and luminosity \citep{Gultekin09, Beifiori12,McConnell13, Lasker14,Reines15}. An exact explanation of this apparently tight coevolution is still missing, although several hypotheses have been proposed. For instance, there might be one specific factor that has a strong influence on both galaxy and its MBH, such as the total gas reservoir (see \citealp[]{Kulier15} and references therein). Another compelling idea is that there is a mutual feedback: galaxy evolution processes regulate the inflow of the material towards its center (thereby the growth of the MBH) and, on the other hand, a rapidly accreting MBH can lead to gas heating and outflows, resulting in suppressed star formation \citep{Maiolino12}.

One of the fundamental problems in the field is related to the masses of MBHs found in the range of $\mathrm{10^4 - 10^{10}~M_{\odot}}$ with the most massive and luminous quasars already being observed at redshifts as high as $z = 7.54$ \citep{Wu15,Mazzucchelli17, Banados18}. Firstly, it is still unclear what type of objects and processes constitute the formation of MBH seeds. There are currently several proposed scenarios, three of which have gained particular interest and are being extensively studied in recent years \citep{Volonteri21}. These include i) low mass seeds ($\mathrm{\sim100~M_{\odot}}$) from population III stars (PopIII), ii) intermediate mass seeds ($\mathrm{\sim 10^4~M_{\odot}}$) from runaway collisions in stellar clusters and iii) heavy seeds ($\mathrm{10^4 - 10^5~M_{\odot}}$) from direct collapse black holes \cite[for an extensive review see][]{Inayoshi20}. The second unknown corresponds to processes responsible for the mass growth and feedback exerted back on the host galaxy. In this case, the main three channels are \citep{Croton06}: i) quasar (radiative) mode characterized by high accretion rates and strong X-ray emission (induced by galaxy mergers and/or disc instabilities) , ii) radio (jet) mode where accretion rates are lower due energy transfer to a jet production and iii) MBH mergers.

It is extremely challenging to test the above models with electromagnetic (EM) observations. This is due to the fact that MBH evolutionary processes started very early in the history of the Universe and the most distant objects that we can now study represent only a small and highly biased fraction of the whole population \citep{Barausse17,Shankar19}. Theoretical predictions indicate that MBH seeds from remnants of PopIII stars could form at redshifts as high as 20 \citep{Volonteri10,Inayoshi20}, while redshifts considered for DCBHs fall approximately between 13 and 20 \citep{Yue14}. This is outside the range of any of current EM observatories, however future missions like SKA in radio \citep{Taylor00}, JWST in IR/optical \citep{Behroozi20}, and Athena in X-rays \citep{Athena13} will probe sources up to redshifts of about 15-20.

Both observations and theory predict that MBHs can form binaries and merge within the Hubble time. It is one of the fundamental assumptions in hierarchical models of galaxy evolution \citep{Benson10} and there is a non-negligible number (a few hundred)  of observed candidate systems that show signs of hosting an inspiraling MBH binary at pc - subpc separations (MBHB; \citealp{DeRosa19, Saade20}). Such systems are expected to emit low-frequency gravitational waves (GWs) in the range of about nHz to mHz, depending on binary mass and separation. As GWs hardly interact with matter, they can carry unperturbed information further than EM radiation and thus may enable us to reach the earliest stages of the Universe's evolution.

LISA (Laser Interferometer Space Antenna, \citealp{Amaro-Seoane17}) is scheduled to start its operation around 2034 and will be sensitive to mergers of MBHBs of around $\mathrm{10^3 - 10^7~M_{\odot}}$, radiating at mHz frequencies up to redshifts well above 20. Moreover, lower (nHz) part of the GW spectrum is the main target for Pulsar Timing Arrays (PTAs) already being in operation: PPTA  \citep{PPTA}, EPTA \citep{EPTA}, and NANOGrav \citep{NANOGRAV}, as well as (most recently)  inPTA \citep{INPTA} and CPTA: \citep{CPTA}. What is particularly important is that PTAs have also already constrained an upper limit for the GW background (GWB) amplitude generated by an assembly of cosmological population of MBHBs to approximately $10^{-14}< A < 10^{-15}$ at a reference frequency of $f = 1/{\rm yr}$ \citep{Shannon15, Babak16, Arzoumanian18}. These measurements now constitute a powerful constraint on theoretical galaxy and MBHBs evolution models, already excluding some of the more optimistic predictions.

In this paper, we present the analysis of MBHB populations generated in the novel semi-analytic galaxy evolution model SHARK \citep{Lagos18} in terms of their detectability with LISA and PTA. The model was successfully tested against standard observational data (i.e., stellar mass function, hydrogen mass function, BH-bulge mass relation, etc.) and was also used in predictions for future missions, namely, SKA and Athena \citep{Amarantidis19}. We note that there has not been any study of the GW emission produced by MBHBs naturally occurring in the hierarchical formation scenario of SHARK and therefore we find it is important to fill in this gap.  
A similar research was conducted by various groups, using both semi-analytic models (SAM; \citealp{Sesana11,Dayal19,Barausse20}) and hydrodynamical simulations (HD; \citealp{Salcido16,Kelley17b, Katz19,Katz20}). Globally, the majority of these works trace a similar set of galaxy and MBHB evolution processes, however, they naturally utilize different models to describe them. Given the insufficiency of observational data and the complexity of discussed systems, we are currently unable to place any constraints that would be strong enough to rule out a substantial number of these models. Detection rates for LISA provided by previous works cover a broad range of several to several hundreds events per year (depending on the assumptions related to MBHB evolution and seeding scenarios), where generally more detections are predicted by SAMs mainly due to larger volumes (see, e.g., discussion in \cite{Katz20}, \cite{Amarantidis19}). Similarly, estimates of the GWB amplitude in the PTA range vary between $10^{-17}< A < 10^{-14}$ \citep{Rosado15}, although most works that include a modeling of MBHB evolution (e.g., due to interactions with galactic environment and orbit eccentricity; \citealp{Kelley17a, Bonetti19,Barausse20}) predict lower amplitudes of $A<10^{-15}$, which are in agreement with observationally constrained upper limits. 
We note that a recent detection of a common spectrum process announced by all three major PTAs (CSP; \citealp[]{Arzoumanian20,Goncharov21,Chen21}) raised new questions and revived ongoing discussions. Despite the fact that the amplitude of the signal is higher than previous limits ($\mathrm{A_{CSP} \sim 2\cdot10^{-15}}$), the astrophysical origin of the signal (primarily as the GWB) is neither confirmed nor ruled out. From a theoretical point of view -- there is no consensus. There are studies which prove that CSP can indeed be the GWB, based on our current knowledge \citep{Middleton21}, but on the other hand, other studies have struggled to find realistic models predicting such high GWB amplitude \citep{Villalba22}. All of the above implies a vital need for further studies.

With this work, we would like to initiate our search for more robust methods of model comparison by utilizing the open-source community code SHARK. In contrast to the majority of previous studies, our methods can be easily reproduced, modified, and updated within a well-built framework. Here, we present our first studies of SHARK capabilities within the GW regime using a few functionalities of the code and simplifying assumptions, which we will extend in a future work.

The outline of the paper is as follows. In Section \ref{sec:2}, we describe SHARK and its physical prescriptions,  particularly those that are important for MBH evolution. We also present the course of our calculations of LISA detectability and the GWB at nHz frequencies. In Section \ref{sec:3}, we show our results based on a set of MBH populations produced by SHARK, including merger rates, mass ratios, GWB, and predictions for the LISA and PTA detection rates. We discuss the results and present our conclusions in Sections \ref{sec:4} and \ref{sec:5}, respectively. The cosmological parameters are taken from \cite{Planck}, specifically $h = 0.6751$, $\Omega_m = 0.3121$, $\Omega_{\lambda} = 0.6891$.


\section{Methods\label{sec:2}}

\subsection{SHARK physics}
Below, we briefly summarize all physical processes that can be currently modeled with SHARK and we refer the reader to \cite{Lagos18} for more details. We place a particular emphasis on the prescriptions directly related to MBHs as the main focus of this paper is to study their impact on low-frequency GW emission (equations are taken directly from \citealp{Lagos18}). 

Before moving to semi-analytic prescriptions, we would like to describe the basis of SHARK, namely, dark matter (DM) halo catalogs and merger trees from N-body simulations.  Notably, SHARK is not restricted to just one particular realization, but instead, it allows the user to evolve galaxies based on any chosen N-body DM catalog. Here, we use L210N1536 simulation from the SURFs suite \citep{Elahi18} with a volume of $\mathrm{   (210~cMpc/h)^3}$  and a spatial resolution of 4.5 ckpc/h (where cMpc and ckpc indicate comoving units). There are $1536^3$ simulated particles, with a single mass of $\mathrm{2.21\cdot 10^8 M_{\odot}/h}$. Currently, the DM halo mass resolution is $\mathrm{10^9~M_{\odot}}$ and galaxy stellar mass resolution is $\mathrm{10^{7}~M_{\odot}}$. Future releases of SURFs will aim at resolving halo and stellar masses down to $\mathrm{10^6~M_{\odot}}$ and $\mathrm{10^4~M_{\odot}}$, respectively, which will be of particular importance for studying poorly constrained coevolution processes in dwarf galaxies.

\subsubsection{Gas cooling\label{subsec:2.1.1}}
Gas cooling is one of the most important processes influencing our results and we tested both of the available models, namely Croton06 \citep{Croton06} and Benson10 \citep{Benson10}. The main difference here is the way of calculating cooling timescales.  \\ \\
$Croton06$ \\
The assumption here is that the cooling timescale $t_{cool}$ is on the order of the dynamical timescale $t_{ff}$, both measured at the cooling radius  ($t_{cool}(r_{cool})/t_{ff}(r_{cool}) = 1$). The cooling rate in the hot halo mode (when the cooling radius is smaller than the virial one) is then given by:
\begin{equation}
    \dot{M}_{cool} = 4\pi\rho_g(r_{cool})r^2_{cool}\dot{r}_{cool}
,\end{equation}
where $\rho_g(r_{cool})$ is the gas density profile. 
\\ \\
$Benson10$\\
In this model, the cooling timescale at the cooling radius is equal to the halo age, given by:
\begin{equation}
    t_{halo} = \frac{\int_{0}^{t} [T_V(t')M_{gas}(t')/t_{cool}(t')] \,dt' }{T_V(t)M_{gas}(t)/t_{cool}(t)}
,\end{equation}
where t is the current time and $T_V$ is the virial temperature.

\subsubsection{Disk instabilities}
The criterion for instability was described in Ostriker\&Peebles (1973) and Efstathiou, Lake \&Negroponte (1982) as:
    \begin{equation}
        \epsilon = \frac{V_{circ}}{\sqrt{1.68GM_{disk}/r_{disk}}}
    ,\end{equation}
where $V_{circ}$ is the maximum circular velocity, $r_{disk}$ is the half-mass disc radius, and $M_{disk}$ is the disc mass (gas plus stars). If $\epsilon < \epsilon_{disk}$, then the disk is considered unstable and $\epsilon_{disk}$ is a free parameter in SHARK.

\subsubsection{Galaxy mergers\label{subsec:2.1.3}}
Orbits of satellite galaxies decay due to dynamical friction with halo material as well as other satellites after entering central galaxy virial radius. This decay time before galaxies merge is calculated following \cite{Lacey93} as:
    \begin{equation}
        \tau_{merge} = f_{df}\Theta_{orb}\tau_{dyn}\left(\frac{0.3722}{ln(\Lambda_{Coulomb})}\right)\frac{M}{M_{sat}}
    ,\end{equation}
where 
\begin{equation}
 \Theta_{orb} = \left(\frac{J}{J_c(E)}\right)^{0.78} \left(\frac{r_c(E)}{R_V}\right) 
,\end{equation}
which is a function of orbital parameters: initial angular momentum (J) and energy (E) of the satellites orbit; $\mathrm{J_c(E)}$ and $\mathrm{r_c(E}$) correspond to angular momentum and radius but for a circular orbit with satellite energy. Then, $\tau_{dyn} = \frac{R_V}{V_V}$ is the dynamical timescale of the halo (for virial radius $R_V$ and velocity $V_V$), $\Lambda_{Coulomb} = ln\left(\frac{M}{M_{sat}}\right)$ is the Coulomb logarithm, where $M$ is the halo mass of the central galaxy and $M_{sat}$ is the mass of the satellite, including the mass of its halo. $f_{df}$ is an adjustable parameter corresponding to approximation of decay time below the resolution of the simulations (here set to 0.1).   

\subsubsection{BH growth and AGN feedback\label{subsec:2.1.4}}
Besides galaxy mergers, the growth of MBHs is assumed to occur during two phases. The first one, the so-called "quasar mode," is characterized by accretion of material during starburst event, which can be triggered either by galaxy merger or disk instabilities. It is a rapid process resulting in high accretion rates and very efficient growth of the MBH. It can be calculated via \citep{Kauffman00}:
    \begin{equation}
        \delta m_{BH,q} = f_{mbh}\frac{m_{gas}}{1+(v_{mbh}/V_{V})^2}
    ,\end{equation}
where $m_{gas}$ and $V_{V}$ are the cold gas mass reservoir of the starburst and the virial velocity, respectively; and $f_{mbh} = 8\cdot10^{-3}$ is responsible for normalization of the BH-bulge relation and $v_{mbh} = 400~km/s$ regulates the binding energy of the system relative to $V_{V}$. \\
Accretion rate is calculated for a bulge accretion timescale ($\tau_{bulge}$):
\begin{equation}
    \dot M_{BH,q} = \frac{\delta m_{BH,q}}{\tau_{bulge}} 
.\end{equation}

The second growth channel is accretion during so called "radio mode" and it can be modeled following Croton16 \citep{Croton16} or Bower06 \citep{Bower06}. In this work, we explore the results for both of these models.

In the case of the Croton16 model, the Bondi-Hoyle like accretion mode is assumed and completed by "maximal cooling flow" model \citep{Nulsen00}, giving:
\begin{equation}
    \dot M_{BH,r} = \kappa_R\frac{15}{16}\pi G\mu m_p\frac{\kappa_BT_{V}}{\Lambda(T_{V})}m_{BH}
,\end{equation}
where $\kappa_R$ is a free parameter regulating accretion rate, $\kappa_B, \Lambda(T_{V})$ are the Boltzmann constant and the cooling function depending on virial temperature ,$T_{V}$, and halo metalicity, $\mu m_p$, is the mean mass per gas particle. MBH luminosity is then given by:
\begin{equation}
    L_{MBH} = \eta \dot M_{BH,r} c^2
,\end{equation}
where $\eta$ is the luminosity efficiency. We explore two model variants with $\eta$ set to either 0.1 (10\% of the infalling matter is converted into radiation) or 0.4 (maximum value for a spinning black hole, 60\% of matter contributes to MBH growth and the rest is emitted as radiation).

In Bower06, cooling depends on the free-fall and cooling radii. If the free-fall radius is much smaller than the cooling radius, then gas is accreted on a free-fall timescale. Otherwise, quasi-hydrostatic cooling is assumed (it is also the case of an effective AGN feedback). For the latter, MBH growth rate is given by: 
\begin{equation}
    \dot M_{BH,r} = \frac{L_{cool}}{0.2c^2}
,\end{equation}
where $L_{cool}$ is the cooling luminosity. It is assumed that an active galactic nucleus (AGN) will quench gas cooling if the available AGN power is related to the cooling luminosity as:
\begin{equation}
    L_{cool} < f_{edd}L_{Edd}
,\end{equation}
where $L_{Edd}$ is the Eddington luminosity of the MBH and $f_{edd}$ is a free parameter in the range of 0.0001 - 0.1. It allows us to manually set the threshold for effective gas quenching as a fraction of Eddington luminosity. We tested two models with $f_{edd}$ equal 0.01 and 0.001, which means that in the latter case, gas cooling will be quenched at lower AGN luminosities.

It is important to note that Croton16 and Bower06 require different approaches to modeling gas cooling. The latter requires an explicit calculation and comparison of cooling and dynamical timescales and so it can be used only with the Benson10 gas cooling model, while Croton16 is used with the Croton06 model (see Section \ref{subsec:2.1.1}).

The major consequence of the above is that in Bower06 models, the AGN feedback will have a stronger dependence on redshift, especially in case of the most massive galaxies (when cooling radius is calculated based on the halo age, it will be shorter at earlier times). This would mean that the higher the redshift, the more probable it is that cooling will occur on free-fall timescales and in this case, AGN feedback is ineffective in Bower06 models \citep{Bower06}. Another important difference to note is that in Cronon16, the AGN feedback depends on gas properties and MBH mass, while in Bower06 it is limited only by the Eddington luminosity \citep{Bower06}.

\subsubsection{Other}
$Star~formation$:\\
We can choose between four different models, which either calculate molecular gas densities from galaxy properties or employ an observationally constrained value. Here, we use the model from \citep{Blitz06} based on observations of nearby galaxies \citep{Leroy13}. \\ \\
$Stellar~and~SNe~feedback:$\\
Here we discriminate two components: gas outflows from galaxy $M_{gal}$ and from halo $M_{halo}$. The rates of the two outflows are always related in the same way, however $M_{gal}$ depends on the supernova (SN) feedback, which can be calculated with six different models. We used \cite{Lagos13}. \\ \\
$Reincorporation~of~ejected~gas:$\\
The ejected gas is added back to the galaxy on a timescale that depends on their masses. This process is fully steerable, mainly by the $\tau_{reinc}$ scaling parameter \citep{Henriques13}. If set to 0, an instantaneous incorporation is implemented. Here, we used $\mathrm{\tau_{reinc} = 25~Gyrs}$.\\ \\
$Photoionization~feedback$\\
In this case, we have two models to choose from: \cite{Lacey16} and \cite{Sobacchi13} -- and we used the latter. The difference between the two is that in \cite{Sobacchi13}, the condition for halo gas cooling is redshift-dependent and employs a fixed value for the UV background $z_{cut} = 10$; however, all model parameters can be adjusted freely. \\ \\
$Environmental~effects$\\
Currently, there are two approaches that are generally implemented to model the halo gas of the satellite galaxies. In the first case, it is assumed that the hot halo gas is stripped from the galaxy as soon as it becomes a satellite (cold gas is maintained). In the second approach, satellite galaxies retain their hot halo gas and continue to use it as long as it is available. In our models, we used the first approach.

\subsection{MBH populations\label{sec:2.2}}

We generated 12 models that vary in three main aspects: 1) halo and MBH seed masses, 2) MBH growth models, and 3) AGN feedback and BH growth parameters. The most relevant simulation parameters and their variations are summarized in Table \ref{table:1}.

For each simulated model, we drew up a list of galaxies that merge within the Hubble time and contain a central MBH. The primary list consists of MBH masses $m1,~m2,$ and redshift, $z,$ of the merger, the secondary list contains detailed information on merging galaxies that will be used to future studies. The percentage of mergers of galaxies both containing a central MBH with respect to all mergers varies between $\mathrm{\sim 8\%}$ and $\mathrm{\sim 60\%}$ for models with most or least massive MBH seed and halo masses, respectively. It is also worth noting that $\sim$85\% of all mergers in all models are minor, namely, they fulfill the condition: $0.1 < M_s/M_p \leq 0.3$, where $M_p$ and $M_s$ are total masses of the primary and secondary galaxy, respectively.
   
\begin{table*}[t]
\caption{Most relevant simulation parameters. The default model parameters are shown in the second column and the third column contains our applied variations. Each MBH growth model was run with three seed mass configurations: B3H9, B4H10, B5H11. Additionally, we check the dependence on two parameters: $\eta$ for Croton16 and $f_{edd}$ for Bower06 models. For a detailed description of these parameters, see Section \ref{subsec:2.1.4}. This gives us a total of 12 models listed explicitly in Table \ref{Table:2}.}            
\label{table:1}      
\centering                      
\begin{tabular}{l l l l}        
\hline                
Parameter & Default & Variation & Description \\
\hline\\     
   MBH growth                   & Croton16     & Bower06         & AGN feedback and BH growth models \\ \\  
   Gas cooling                  & Croton06     & Benson10        & Gas cooling timescales \\ \\  
    $\mathrm{m_{seed}~[M_{\odot}]}$          & $10^4$       & $10^3$, $10^{5}$   & MBH seed masses (marked as B3 - $\mathrm{10^3~M_{\odot}}$, B4 - $\mathrm{10^4~M_{\odot}}$, B5 - $\mathrm{10^5~M_{\odot}}$) \\ \\      
   $\mathrm{m_{seed}^{halo}~[M_{\odot}]}$   & $10^{10}$    & $10^{9}$, $10^{11}$ & Halo seed masses (marked as H9 - $\mathrm{10^9~M_{\odot}}$, H10 - $\mathrm{10^{10}~M_{\odot}}$, H11 - $\mathrm{10^{11}~M_{\odot}}$)\\ \\
   $\mathrm{f_{edd}}$           & 0.01         & 0.001           & Eddington luminosity scaling factor (only for Bower06) \\ \\
   $\mathrm{\eta}$              & 0.1          & 0.4             & Luminosity efficiency dependent on BH spin (only for Croton16) \\ \\
  
\hline
\end{tabular}
\end{table*}

\subsection{Assumptions for the GW emission}
The current version of SHARK does not trace the binary evolution of MBHs, which simply merge immediately after the merger of their host galaxies (we note that this time is different from the DM halo merger time because of a delay due to dynamical friction, as explained in Section \ref{subsec:2.1.3}). In our work we did not include any kind of post-processing (apart from a few assumptions) and we tested performance of SHARK "as is," therefore results presented here should be considered as upper limits. Moreover, we will use these results as our control sample for further studies which will be presented elsewhere.

In order to make the first predictions for GW detectors, we have to either calculate or assume specific timescales upon which the emission is produced (it is related to the orbital frequency evolution with time - the information which we are missing from the simulations). MBHBs that could be detected with LISA will fall in the mass range of about $10^3 - 10^7~M_{\odot}$, which corresponds to merger times (with respect to mission lifetime) from days to years. Calculation of an expected signal-to-noise ratio (S/N) requires these times to be specified, and in most recent studies, they are drawn from a uniform distribution of 0-4 years (marking the nominal length of LISA's operation). A more realistic approach would be to also include sources merging after LISA observations but for which we could still detect the inspiral phase. We therefore use a wider distribution of merger times (see Section \ref{subsec:2.3.1}) limited mostly by computational expenses. We note however, that the true statistical relevance of a given merger time distribution would be revealed for models including MBHB evolution processes.

 Gravitational wave background, which is a primary target of worldwide PTA efforts, is expected to be dominated by signals from the most massive MBHBs ($M>10^8~M_{\odot}$), for which merger times could be on the order of billions of years. In the first instance, the amplitude of the GWB would depend on galaxy merger rates and the number density of MBHBs, which is virtually unknown due to the lack of observational data. Furthermore, there is a number of processes which will further modify the spectral shape and strain of the signal that are mostly dependent on the relations between MBHs and their hosts. These include growth mechanisms, binary formation channels, and orbital separation hardening (see e.g., \citealp{Kelley17a, Kelley17b}). In our study we do not address these issues directly and instead assume a fixed time delay between the merger of host galaxies and their MBHs (see Section \ref{subsec:2.3.2}). This allows us to focus entirely on the influence of different growth and feedback models and test if current version of SHARK is in agreement with PTA observational constraints.

\subsubsection{LISA\label{subsec:2.3.1}}
We calculated the S/N for LISA four-year long mission using $gwsnrcalc$ Python package included in the BOWIE analysis tool \citep{Katz19}, which uses PhenomD for waveform generation \citep{Husa16, Khan16}. The advantage of the code is that it enables us to perform calculations for all phases separately (inspiral, merger, rigndown), as well as including BH spins and taking into account binaries that may merge outside LISA's observational window.
In our calculations, we follow a similar procedure as in \cite{Katz20}: we found the number of binaries merging within 50 years using Eq. \ref{eq:13}  (without any assumptions on the S/N) and randomly chose a list of our sources from the given distribution, assigning to each of them a merger time from 0 to 50 years (with 1 year step) and start-end times of the event with respect to LISA operations. We leave a detailed spin analysis for future works and thus we set a constant value for all MBHs here, depending on the chosen luminosity efficiency $\eta$ (for $\eta = 0.1$ we set $a = 0.1$ and $\eta = 0.4$ would correspond to maximally spinning BH with $a = 1$). \\
The averaged S/N is calculated via \citep{Robson19}:
\begin{equation}
     \langle \rho^2 \rangle = \frac{16}{5} \int_{0}^{\infty} \frac{h_c^2}{h_N^2}\frac{1}{f} \,df
,\end{equation}
where $h_c$ and $h_N$ are the characteristic strains of the gravitational wave and detector sensitivity, respectively. 

We set a detection S/N threshold to 7. The expected number of mergers per year is then calculated via \citep{Arun09}:
\begin{equation}
 \label{eq:13}
    \frac{d^2N}{dzdt} = 4\pi c N_{com}(z)\left(\frac{d_L(z)}{1+z}\right)^2~ [year^{-1}],
\end{equation}
where $d_L(z)$ is the luminosity distance, c is the speed of light, and
\begin{equation}
 \label{eq:11}
    N_{com}(z) = \frac{d^2n(z)}{dzdV_c} \approx \frac{N(z)}{\Delta zV_c}
\end{equation}
is the BH merger rate density (per unit comoving volume per redshift interval) and $V_c$ is the comoving volume of our simulations \citep{Katz19}.

\subsubsection{PTA\label{subsec:2.3.2}}
In order to characterize the GWB from the whole population of MBHs for each of our models, we start by calculating the characteristic strain $h_c$ following \cite{Rosado15}. Here, we limited our analysis to binaries with total masses $M\geq 10^8~M_{\odot}$ and $z\leq 2$ because we expect only the most massive and relatively close sources to contribute to GWB at nHz frequencies available for PTA. All of these sources were then assigned with two parameters drawn from uniform distributions, namely: $\iota$ (binary inclination with respect to the line of sight) and $t$ (estimated time to merger). In case of the latter, we probed two scenarios, with $t$ from 0 to either 100 Myr or 1 Gyr.

The strength of GWB can be measured in terms of characteristic strain, which is an incoherent superposition of strain from each binary as a function of frequency:
\begin{equation}
        h_c^2 = \frac{\sum_k{h_k^2f_k}}{\Delta f}
    ,\end{equation}
where summation is running over all sources at a given frequency bin determined by the whole observation time $T$ of an array of pulsars ($\Delta f = 1/T$). The average strain for each binary is given by:
\begin{equation}
    h_k = A\sqrt{\frac{1}{2}(a^2+b^2)}
,\end{equation}
where
\begin{equation}
A = 2\frac{G^{5/3}\mathcal{M}^{5/3}(\pi f)^{2/3}}{c^4r}
\end{equation}
is the dimensionless amplitude of the signal, $f$ is the rest frame frequency, and
\begin{equation}
    \begin{split}
        a &= 1+\cos^2(\iota)\\
        b & = -2\cos(\iota)
    \end{split}
,\end{equation}
which are two contributions from wave polarizations. The comoving distance $r$ is defined as: 
\begin{equation}
    r = \frac{c}{H_{100}h}\int_{0}^{z} (\Omega_m(1+z)^3+\Omega_{\lambda})^{-1/2} \,dz.
\end{equation}
Given the time to merger assigned to every binary, we calculate the frequency of its GW emission via:
\begin{equation}
       f = \left(\frac{8}{3}(\pi)^{8/3}\frac{96}{5}\frac{G^{5/3}}{c^5}\mathcal{M}^{5/3}t\right)^{-3/8}.
\end{equation}


\section{Results\label{sec:3}}

\cite{Lagos18} presented a detailed analysis of a model, which is marked as B4H10-n1 here (BH seed mass of $\mathrm{10^4~M_{\odot}}$ and halo seed mass $\mathrm{10^{10}~M_{\odot}}$, Croton16 model with $\mathrm{\eta = 0.1}$). Additionally, we extend the analysis to a wider set of MBH/halo seed masses and two different models of MBH growth as summarized in Table \ref{table:1}. 

\begin{figure}[h!]
\centering
    \resizebox{\hsize}{!}{\includegraphics{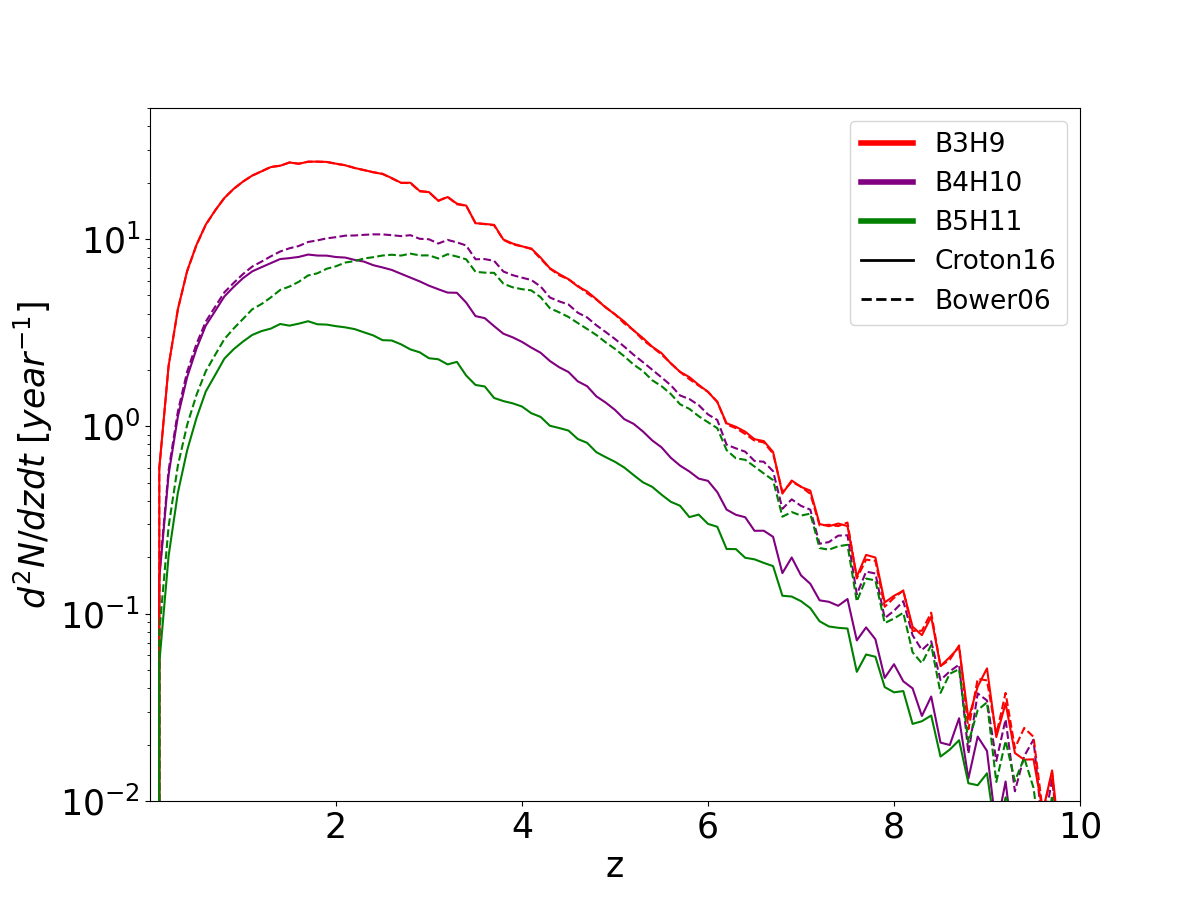}}
    \caption{Merger rates per year for Croton16 and Bower06 models including all galaxies with a central MBH. The plot shows the basis MBH growth models (with $\eta = 0.1$ and $f_{edd} = 0.01$) as variations in these parameters does not affect the result.}
    \label{Fig1}
\end{figure}

Figure \ref{Fig1} shows merger rates per year calculated with Eq. \ref{eq:13} for both MBH growth models (Croton16 and Bower06) and three seed scenarios. The plot does not show models with changed $\eta$ and $f_{edd}$ parameters as they overlap with the default realizations (they do not affect the merger rates). It can be seen that the highest merger rate for Bower06 models is shifted towards higher redshifts with respect to Croton16 models. Moreover, the shift is growing bigger with increasing seed masses, however the Croton16 and Bower06 models are indistinguishable for the lightest seed scenario (B3H9 plotted with red color). This comes from the fact that, as noted in \cite{Croton06, Bower06}, the cooling timescale is irrelevant for small halos. Additionally, we could expect that if the two considered models use distinct relations between cooling and dynamical radii (see Section \ref{subsec:2.3.1}), the results will diverge for larger halos. Lastly, Croton16 models (solid) show a slightly larger difference (approximately by a factor of 1.5) in merger rates between the medium (purple) and highest (green) seed masses than Bower06 models (dashed).

Below, we present the results of calculating low frequency GW emission from inspiraling binaries for all 12 MBH populations. In Section \ref{sec:3.1}, we focus on mHz band that will be probed by LISA and in Section \ref{sec:3.2}, we show the predicted GWB at nHz frequencies which is the main target of PTA observations.

\subsection{LISA\label{sec:3.1}}
Table \ref{Table:2} contains the list of all 12 models analysed in this work along with merger and detection rates per year for the LISA mission. In the third column, we report the percentage of mergers where both galaxies contain a central MBH with respect to all mergers. As seen in Figure \ref{Fig1}, the total number of mergers per year is higher for Bower06 models (except the lightest seed scenarios). This results in approximately ten more detections with respect to Croton16 models for each corresponding seed scenario. 

In general, detection rates vary between $\sim 60$ to $\sim 7$ per year for the least and most massive seed models, respectively. In the case of the variable $\eta$ parameter for Croton16 models, the ones with $\eta = 0.4$ result in a few more detections. This mostly comes from the fact that GW signal detected by LISA will be higher for systems with larger spins. In case of Bower06 models with varying $f_{edd}$, we do not find any significant differences in detection rates which means that they are purely statistical and originate from our sampling procedure.

\begin{figure}[h]
\centering
    \resizebox{\hsize}{!}{\includegraphics{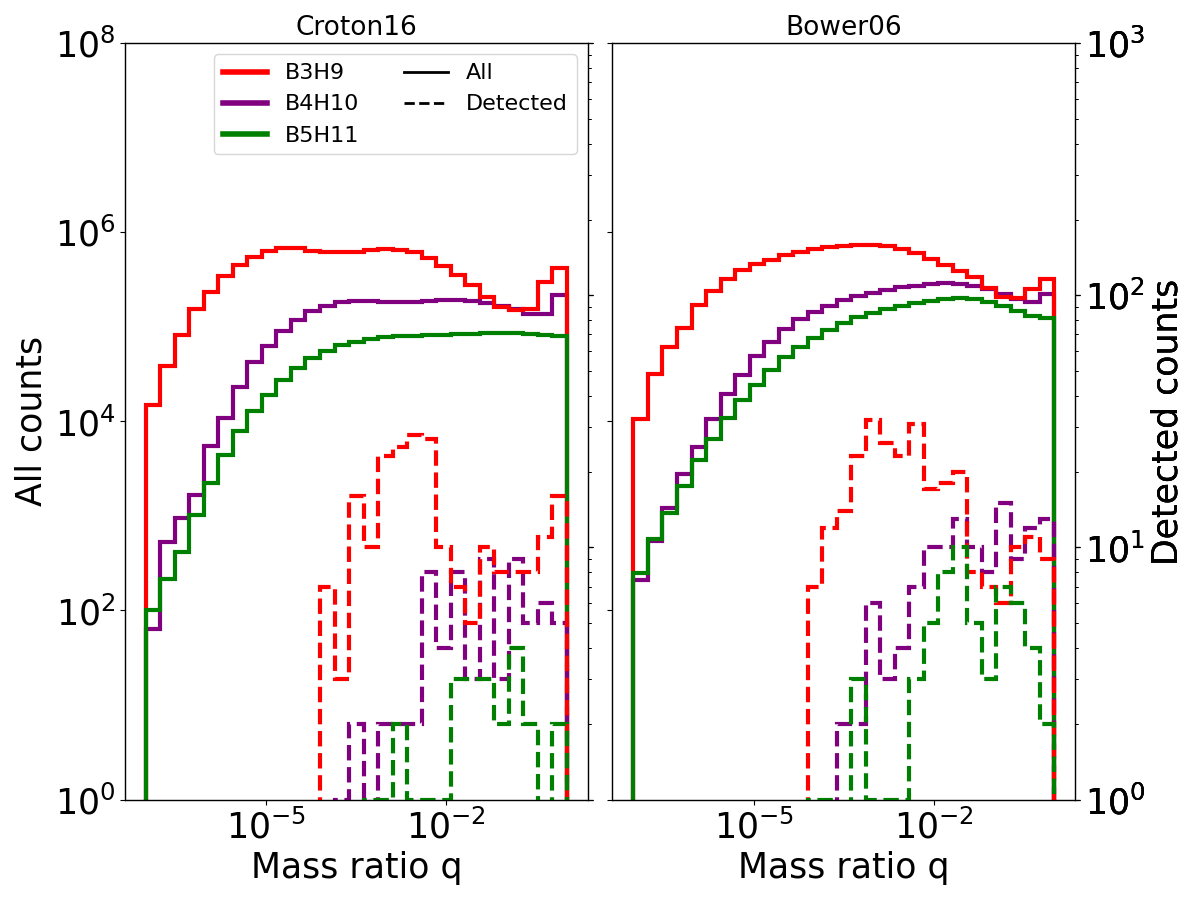}}
    \caption{Mass ratios for all binaries generated in SHARK (solid line) and those detected over four years by LISA (dashed line) for default Croton16 and Bower06 models, i.e., with $\eta=0.1$ and $f_{edd} = 0.01$.}
    \label{Fig2}
\end{figure}

Figure \ref{Fig2} shows mass ratios for all binaries (solid lines) and those detected within four years of LISA operation (dashed lines). Here, we also report only the default versions of both Croton16 and Bower06 models. From the plot we can see that both models (in all seed scenarios) favor binaries with low mass ratios ($q < 10^{-2}$). Particularly, there is a depletion of binaries with $q$ in the range of $10^{-2} - 10^{-1}$ for the least massive seed scenario (red color) however it's less significant for Bower06 models. This results in a much smaller detection rate in this $q$ regime. For higher mass seeds, Croton16 models show a plateau, producing a relatively even sample of binaries with various mass ratios above $q = 10^{-4}$, while in case of Bower06 models, the counts are steadily growing towards higher mass ratios and slightly dropping down at the end where $q = 1$. As noted in Section \ref{sec:2.2}, the vast majority of mergers are minor (irrespectively of the model), which explains the distribution of $q$ in Figure \ref{Fig2}.

Comparing Figures \ref{Fig1} and \ref{Fig2} with information from Table \ref{Table:2}, we can see that higher detection rates in case of Bower06 models may come from two main factors: a) these models are characterized with higher merger rates (there are more halos containing a MBH); and b) their binary population consists of systems with higher mass ratios (producing stronger signal in the detector). This would mean, that apart from being more effective in seeding halos with MBHs (due to directly comparing cooling and dynamical radii; see Sections \ref{subsec:2.1.1} and \ref{subsec:2.1.4}), the Bower06 model results in a more even MBH mass distribution. We find that there are approximately 19\% and 45\% more mergers with $q>10^{-2}$ for Bower06 medium and high seed scenarios, respectively.

\begin{table}[h]
\caption{Merger and detection rates per year calculated for the LISA mission. In case of Croton16, the $n1$ and $n2$ refer to models with $\eta$ set to 0.1 or 0.4, respectively. For Bower06, $f1$ and $f2$ correspond to $f_{edd} = 0.01$ and $f_{edd} = 0.001$, respectively. The fourth column shows the percentage of mergers of galaxies both containing central MBHs with respect to all mergers.}
\label{Table:2}  
\begin{tabular}{l c c c c}      
\hline        \\         
   Model            & Detected  & Total & w/ 2 MBHs & Version  \\ \\ 
\hline
Croton16\\
\hline \\
    B3H9-n1          & 55.5     & 84.32  & 62.3\%    & \\
    B4H10-n1         & 17.25    & 26.42  & 17.5\%    & $\eta = 0.1$ \\
    B5H11-n1         & 7.0      & 11.73  & 7.8\%     & \\ \\

    B3H9-n2          & 62      & 84.37     & 62.3\%    & \\
    B4H10-n2         & 22.5   & 26.44     & 18.3\%    & $\eta = 0.4$\\
    B5H11-n2         & 11.25  & 11.73     & 7.8\%     & \\ \\

\hline
Bower06\\
\hline \\
    B3H9-f1          & 68.5  & 84.17    & 62.2\%    & \\
    B4H10-f1         & 31  & 40.01    & 22.5\%    &  $f_{edd} = 0.01$ \\ 
    B5H11-f1         & 15.25  & 30.37    & 14.5\%    &\\ \\

    B3H9-f2          & 64.95  & 84.16    & 62.2\%    & \\
    B4H10-f2         & 31     & 40.01    & 22.5\%    & $f_{edd} = 0.001$\\
    B5H11-f2         & 22     &  30.36   & 14.5\%    & \\ \\
\hline
\end{tabular}
\end{table}

\subsection{PTA\label{sec:3.2}}
\begin{figure*}[h!]
\centering
    \includegraphics[width=17cm]{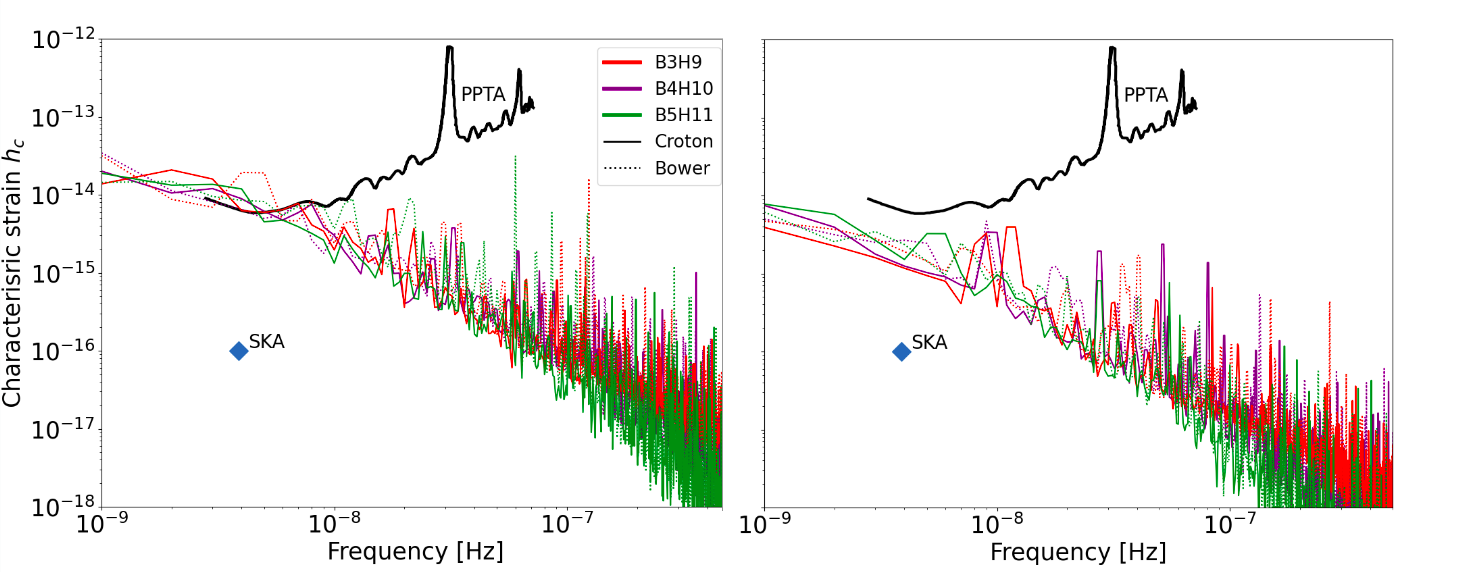}
    \caption{GWB characteristic strain comparison between default Croton16 and Bower06 models ($\eta = 0.1$ and $f_{edd} = 0.01$). Panel on the left shows models with time to coalescence up to 100 Myr, while on the right, the maximum time is 1 Gyr. The black line and blue diamond show the observationally constrained limit (PPTA; \citealp{Shannon15}) and estimated SKA sensitivity (at the most most sensitive frequency), respectively.}
    \label{Fig3}
\end{figure*}

In terms of the nHz GWB, it is practically impossible to distinguish between various seed scenarios; in addition, the two BH growth models result in a comparable signal amplitude, which is shown in Figure \ref{Fig3}. All Bower06 models produce a slightly higher characteristic strain due to the same two factors already mentioned for LISA case (higher merger rates and mass ratios), but the difference is miniscule and thus not statistically relevant. As GWB is produced by only the most massive ($\mathrm{M > 10^8~M_{\odot}}$) and closest ($\mathrm{z<2}$) binaries and given that MBH evolution and growth are partially driven by mergers, the memory of the seeding scenario is lost. 

However, GWB might carry important information on MBH binary evolution after the galaxy merger. This regime is particularly intriguing as it's poorly understood due to our limited resolution in both theory and observations. As we mentioned before, in this work, we did not focus on binary evolution and instead we qualitatively showed how the time available for binary formation and evolution affects the predicted GWB strain amplitude. 

Assuming an optimistic scenario, where all galaxy mergers result in MBH binary formation and subsequently coalescence within 100 Myrs we obtain a mean characteristic strain amplitude at the reference frequency ($f = 1/yr$) $h_{c,yr} = 1.087\cdot 10^{-15}$ averaged over all models. When the coalescence time is additionally delayed up to 1 Gyr, the average strain amplitude drops down to $h_{c,yr} = 1.376\cdot10^{-16}$. Both scenarios are shown in two panels of Figure \ref{Fig3} (shorter coalescence time on the left plot and longer time on the right). 

What is extremely important when considering this scenario is that current PTA observations already put rigorous constraints on possible GWB -- we compare them with our results directly in Table \ref{table:3}. Additionally, both panels of Figure \ref{Fig3} show sensitivity curve of PPTA \citep{Shannon15} for better visibility. These results can lead to a few conclusions. First, we can see that both coalescence times we applied fit within observational constraints, however, GWB amplitude for the 100 Myr scenario is close to currently achievable sensitivities. On the other hand, a recently detected CSP has a slightly higher (on average by a factor of $\sim2$) amplitude which is also summarized in Table \ref{table:3}. Notably, this apparent discrepancy may have been caused by imperfect solar system ephemeris models used in previous studies (for a discussion see \citealp[]{Arzoumanian20}). Nevertheless, our predicted amplitudes lie below the observed CSP, which would disfavor the intensely debated hypothesis that the signal  is indeed the GWB.
The second conclusion is that increasing the delay time between galaxy and MBH merger by an order or magnitude decreases the characteristic strain amplitude by approximately half an order of magnitude. Moreover, the delay time is the only parameter that significantly affects the resultant GWB in our studies. A further decrease in the amplitude may be caused by additional processes, which we have omitted here,  for instance, recoil kicks and ejections (lowering the number density of MBHBs), binaries stalling at $\mathrm{\sim~kpc}$ separations and never reaching the GW emission regime or longer delay times caused by environmental coupling. Here, we show that even considering our optimistic upper limits, current PTAs are still incapable of detecting the GWB. We note however, that given fast advances in instrumentation and analysis (along with ever increasing timing baseline), our sensitivities are increasing by approximately half an order of magnitude with every new data release (see e.g., \citealp[]{Aggarwal19}). What is more, future detectors like SKA are expected to reach at least two orders of magnitude lower compared to today's experiments (shown in Figure \ref{Fig3} with blue diamond) and will have a significant impact on the detection and characterization of GWB.

\begin{table}[h]
\footnotesize
\caption{Summary of observational upper limits (UP) and common spectrum process (CSP) amplitudes from PPTA \citep{Shannon15,Goncharov21}, NANOGrav \citep{Arzoumanian18,Arzoumanian20}, and EPTA \citep{Babak16,Chen21}. Last column shows mean amplitudes averaged over all SHARK models (range corresponds to 100 Myr and 1 Gyr coalescence times). Each value is referenced to the frequency of $ 1/{\rm yr}$.}
\label{table:3}  
\begin{tabular}{c c c c c}      
\hline\\               
             & PPTA               & NANOGrav            & EPTA               & SHARK  \\ \\
\hline \\
UP           & $1.0\cdot10^{-15}$   & $1.45\cdot10^{-15}$ & $<10^{-14}$   &  \\ \\
             &                    &                     &                       & $10^{-16}-10^{-15}$ \\\\
CSP          & $2.2\cdot10^{-15}$ & $1.9\cdot10^{-15}$  & $2.95\cdot10^{-15}$ &  \\ \\
\hline
\end{tabular}
\end{table}




\section{Discussion\label{sec:4}}

\begin{table*}[h]
\centering
\caption{LISA detection rates obtained by several other groups and methods. In \cite{Dayal19} rates are calculated only for mergers at $ z> 4$. \cite{Bonetti19} and \cite{Barausse20} use the same formalism for seed formation, i.e., defining the instability of protogalactic disks given the Toomre parameter $Q_C$, but using a different threshold value ($Q_C$ equal to either 2.5 or 3, respectively, where a higher value corresponds to more seeds being produced). Binary evolution models: DF (dynamical friction), SH (stellar hardening), G (gas induced migration), Tri (triple interactions), GAL (galaxy gas properties), GW (gravitational waves). The third column lists types of used DM merger trees: ePS (Extended Press-Schechter algorithm; \citealp[]{Press74,Parkinson08}, EAGLE \citealp[]{Crain15,Schaye15}, Illustris \citealp{Illustris}). Each cited work uses a different set of prescriptions for BH growth and feedback, but we do not include all references in the summary as such comparison is beyond the scope of this paper.}
\label{table:4}  
\begin{tabular}{l l l l l l}      
\hline \\               
Reference            & Type     & Basis     & Seeds [$\mathrm{M_{\odot}}$]   & Binary evolution  & Detection rate [1/yr] \\ \\
\hline \\
This work            & SAM      & SURFs    & $10^{3-5}$             & None         & $\sim 7 - 68$ \\ 
\cite{Barausse20}    & SAM      & ePS       & $300$, $10^5$         & DF, SH, G, Tri, GW             & $\sim 4 - 322$ \\
\cite{Dayal19}       & SAM      & SAM       & $150$, $10^{3-5}$     & None         & $\sim 3 - 6$ \\
\cite{Sesana11}      & SAM      & ePS       & $100$, $10^4$         & None         & $\sim 25$ \\
\cite{Bonetti19}     & SAM      & ePS       & $300$, $10^5$         & DF, SH, G, Tri, GW  & $\sim 12 - 75$ \\
\cite{Salcido16}     & HD       & EAGLE     & $10^{4-5}$            & GAL        & $\sim 2$ \\
\cite{Katz20}        & HD       & Illustris & $10^5$                & DF, SH, G, GW  & $\sim 0.5 - 1$ \\

\hline
\end{tabular}
\end{table*}

\begin{figure}[h]
\centering
    \resizebox{\hsize}{!}{\includegraphics{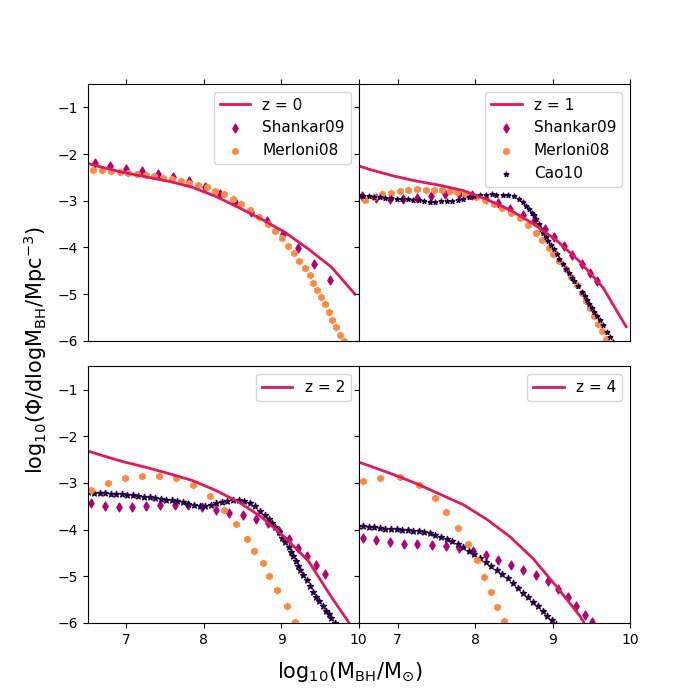}}
    \caption{Black hole mass function (BHMF) for z = 0, 1, 2, 4 from Croton16 model B4H10 (pink solid line). Observationally constrained mass functions are taken from \cite{Merloni08} (orange circles, here z = 0.1 instead of 0), \cite{Shankar09} (purple diamonds), and \cite{Cao10} (dark blue stars).}
    \label{Fig4}
\end{figure}

To gain more insight into our MBH population properties, in Figure \ref{Fig4} we show a BH mass function (BHMF) for a default Croton16 model at four redshift values (z = 0, 1, 2, 4) and compare it with three different observational estimates. It can be seen that the mass function is reproduced reasonably well at the lowest redshift (with a mild overestimation for $\mathrm{M_{MBH}>10^9~M_{\odot}}$) but, generally,  there is a clear discrepancy between the model and observations for the least and most massive MBHs at higher redshifts. This is a common feature  with regard to the majority of simulations discussed in the literature (see e.g., \citealp[]{Habouzit20}) and the reason can be twofold. First, all of the simulations are limited by resolution and volume which directly impact their ability to reproduce the lowest and highest mass BH populations. The second issue can be noticed immediately when looking at Figure \ref{Fig4}. As observational BH mass estimates are mostly limited to the local Universe, deriving BHMF evolution with redshift requires adopting some assumptions or models, e.g. various MBH-galaxy scaling relations (for an extensive review see \citealp{Kelly12}). As a result each estimated BHMF will be different, especially at highest redshifts and any comparison with simulations should be done with caution.

\begin{figure}[h]
\centering
    \resizebox{\hsize}{!}{\includegraphics{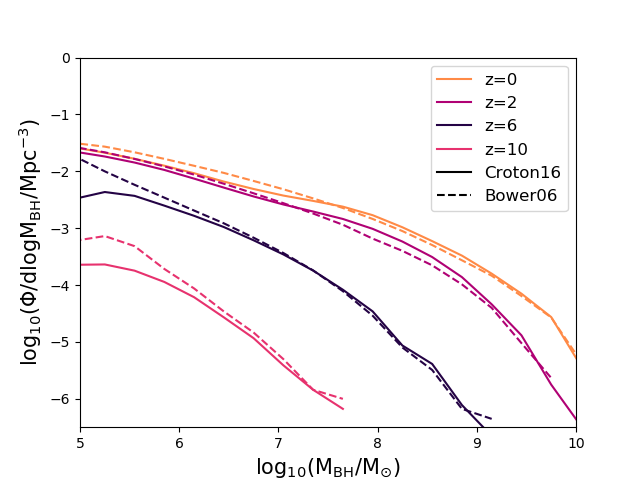}}
    \caption{BHMF comparison between Croton16 and Bower06 models at z = 0, 2, 6, 10.}
    \label{Fig5}
\end{figure}

The BH mass functions for different seed scenarios and BH growth models show little difference and this could mean that the global evolution of mass assembly is similar for all of the models (shown in Figure \ref{Fig5}). This is due to the fact that all models are calibrated with the same $\mathrm{M_{MBH}-M_{bulge}}$ relation, which regulates the rate of gas inflow onto the MBH during a starburst event (quasar mode, for more, see Section \ref{subsec:2.1.4}), which is also the main channel of MBH growth for all SHARK models. However, the BHMF for the two models is clearly different at the lowest mass range ($\mathrm{M<10^6~M_{\odot}}$) and at large redshifts ($z>6$). This could mean that variances in gas cooling prescriptions affecting the second channel of MBH growth (radio mode) mainly influence the seed formation stage; namely, Bower06 predicts more MBH seeds planted into DM halos.

\begin{figure*}[h]
\centering
    \includegraphics[width=17cm]{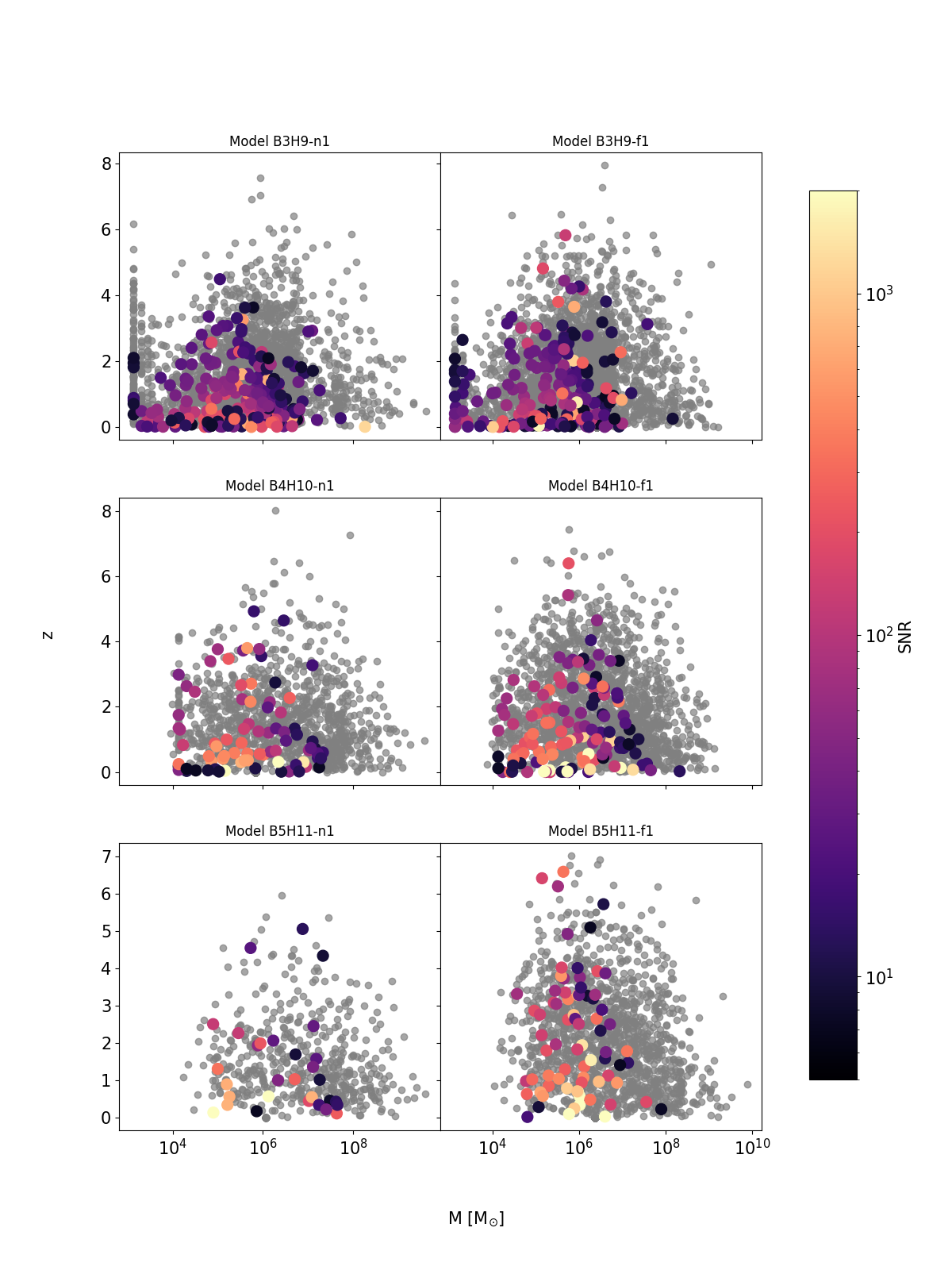}
    \caption{Total masses of binaries detected by LISA over the whole four-year long mission (color) and all binaries merging in 50 years (grey) as a function of redshift. Color scale shows predicted S/N of detection. Croton16 models are shown on the left side, while Bower06 on the right. MBH and halo seed masses are increasing from top to bottom.}
    \label{Fig6}
\end{figure*}

Finally, in Figure \ref{Fig6}, we show total masses of binaries detected by LISA as a function of redshift (color points) overplotted on all binaries merging within 50 years (grey points). It can be clearly seen how the difference between Croton16 and Bower06 models grows with increasing MBH seed mass. In case of Bower06, mergers start occuring at earlier times then for Croton16 for all seed scenarios and the majority of detections comprises binaries with masses in the range of $\mathrm{10^5 - 10^7~M_{\odot}}$. The most distant detected mergers occur at redshifts $z = 6 - 7$ for Bower06 high seed scenario models. Notably, the mass - redshift distribution of detections is very sensitive to model variations and could be used to make preliminary evolutionary constraints.

\section{Conclusions\label{sec:5}}
We present an analysis of 12 MBHB populations generated with SHARK in terms of their detectability with current and future gravitational wave detectors. Specifically, we focused on merger and detection rates for the LISA mission and the predicted GWB amplitude at nHz frequencies probed by PTA experiments. 

Our findings can be summarized as follows:
\begin{enumerate}
    \item GWB amplitude mainly depends on the delay time between host galaxy and MBH merger and is rather independent of all other parameters we investigated (different seed scenarios and growth and feedback models). The GWB spectral shape will depend on the mass spectrum, orbit eccentricity or environmental effects, however, we will investigate this in more detail in future work. Here, we assumed all binary inspirals to be circular and GW driven for simplicity. 
    \item The predicted GWB for all discussed models and the delay time of 100 Myrs is close to the detection limit when compared with the PTA observational constraints. However, both 100 Myr and 1 Gyr delay scenarios lie well below the amplitude of common-spectrum process detected by \cite{Arzoumanian20,Goncharov21,Chen21}.
    Given that our models describe an optimistic scenario, where all MBHs form bound binaries immediately after galaxy mergers and no remnants are expelled from bulges, we can conclude that our studies disfavor the hypothesis of CSP indeed being the GWB.
    \item All models are in a fairly good agreement with the observationally constrained black hole mass functions (BHMF), however, we note some discrepancies for the lowest and highest mass range at redshifts of $z>1$. We note, however, that such deviations are common among most simulations (due to limited resolution and volume; e.g., \citealp[]{Habouzit20}) and also that different observational estimates vary substantially, especially for high redshifts (due to a lack of data) and thus any comparisons should be made with caution. 
    \item Table \ref{table:4} summarizes the key features of our and previous studies. In particular, we note types of merger trees, seeding scenarios, MBHB evolution models, and LISA detection rates. Despite the substantial differences at all levels of modeling, estimated merger rates are rather consistent (e.g., due to the parameter calibration with respect to the observed scaling relations) and therefore it could be difficult to evaluate the accuracy of each model. One way to robustly estimate the contribution of each model to the detection rates would be to perform large-scale comparative studies within one consistent framework, such as SHARK. This is the undertaking we have initiated with the present work. We showed that the mass-redshift distribution of the detected sources strongly depends on the evolution models (of both MBH seeds and their growth) and given our uniform framework, this should allow us to place new constraints on galaxy evolution when compared to the actual detections we expect in the future.
    \item Finally, we also find that the mass ratios of the majority of detectable binaries will be far from unity, implying the importance of developing analysis techniques for high and extreme mass ratio inspirals. 
\end{enumerate}

The presented results should be considered as the upper limits given the assumptions applied in our studies, namely: all galaxy mergers result in formation of a MBHB, preconceived MBH seed masses, and merger delay times. Future works will extend our analysis to the details of MBH binary evolution based on the characteristics of merging galaxies (specifically related to the problem of merger time delays) and the possible electromagnetic signatures of MBH binary inspirals. 


\begin{acknowledgements} TB was suported by the Foundation for Polish Science grant TEAM/2016-3/19. We are grateful to the SHARK team for helpful discussions.   \end{acknowledgements}

\bibliographystyle{aa}
\bibliography{bib2}

\begin{thebibliography}{81}
\expandafter\ifx\csname natexlab\endcsname\relax\def\natexlab#1{#1}\fi

\bibitem[{{Aggarwal} {et~al.}(2019){Aggarwal}, {Arzoumanian}, {Baker},
  {Brazier}, {Brinson}, {Brook}, {Burke-Spolaor}, {Chatterjee}, {Cordes},
  {Cornish}, {Crawford}, {Crowter}, {Cromartie}, {DeCesar}, {Demorest},
  {Dolch}, {Ellis}, {Ferdman}, {Ferrara}, {Fonseca}, {Garver-Daniels},
  {Gentile}, {Hazboun}, {Holgado}, {Huerta}, {Islo}, {Jennings}, {Jones},
  {Jones}, {Kaiser}, {Kaplan}, {Kelley}, {Key}, {Lam}, {Lazio}, {Levin},
  {Lorimer}, {Luo}, {Lynch}, {Madison}, {McLaughlin}, {McWilliams},
  {Mingarelli}, {Ng}, {Nice}, {Pennucci}, {Pol}, {Ransom}, {Ray}, {Siemens},
  {Simon}, {Spiewak}, {Stairs}, {Stinebring}, {Stovall}, {Swiggum}, {Taylor},
  {Turner}, {Vallisneri}, {van Haasteren}, {Vigeland}, {Witt}, {Zhu}, \&
  {NANOGrav Collaboration}}]{Aggarwal19}
{Aggarwal}, K., {Arzoumanian}, Z., {Baker}, P.~T., {et~al.} 2019, \apj, 880,
  116

\bibitem[{{Amarantidis} {et~al.}(2019){Amarantidis}, {Afonso}, {Messias},
  {Henriques}, {Griffin}, {Lacey}, {Lagos}, {Gonzalez-Perez}, {Dubois},
  {Volonteri}, {Matute}, {Pappalardo}, {Qin}, {Chary}, \&
  {Norris}}]{Amarantidis19}
{Amarantidis}, S., {Afonso}, J., {Messias}, H., {et~al.} 2019, \mnras, 485,
  2694

\bibitem[{{Amaro-Seoane} {et~al.}(2017){Amaro-Seoane}, {Audley}, {Babak},
  {Baker}, {Barausse}, {Bender}, {Berti}, {Binetruy}, {Born}, {Bortoluzzi},
  {Camp}, {Caprini}, {Cardoso}, {Colpi}, {Conklin}, {Cornish}, {Cutler},
  {Danzmann}, {Dolesi}, {Ferraioli}, {Ferroni}, {Fitzsimons}, {Gair}, {Gesa
  Bote}, {Giardini}, {Gibert}, {Grimani}, {Halloin}, {Heinzel}, {Hertog},
  {Hewitson}, {Holley-Bockelmann}, {Hollington}, {Hueller}, {Inchauspe},
  {Jetzer}, {Karnesis}, {Killow}, {Klein}, {Klipstein}, {Korsakova}, {Larson},
  {Livas}, {Lloro}, {Man}, {Mance}, {Martino}, {Mateos}, {McKenzie},
  {McWilliams}, {Miller}, {Mueller}, {Nardini}, {Nelemans}, {Nofrarias},
  {Petiteau}, {Pivato}, {Plagnol}, {Porter}, {Reiche}, {Robertson},
  {Robertson}, {Rossi}, {Russano}, {Schutz}, {Sesana}, {Shoemaker}, {Slutsky},
  {Sopuerta}, {Sumner}, {Tamanini}, {Thorpe}, {Troebs}, {Vallisneri},
  {Vecchio}, {Vetrugno}, {Vitale}, {Volonteri}, {Wanner}, {Ward}, {Wass},
  {Weber}, {Ziemer}, \& {Zweifel}}]{Amaro-Seoane17}
{Amaro-Seoane}, P., {Audley}, H., {Babak}, S., {et~al.} 2017, arXiv e-prints,
  arXiv:1702.00786

\bibitem[{{Arun} {et~al.}(2009){Arun}, {Babak}, {Berti}, {Cornish}, {Cutler},
  {Gair}, {Hughes}, {Iyer}, {Lang}, {Mandel}, {Porter}, {Sathyaprakash},
  {Sinha}, {Sintes}, {Trias}, {Van Den Broeck}, \& {Volonteri}}]{Arun09}
{Arun}, K.~G., {Babak}, S., {Berti}, E., {et~al.} 2009, Classical and Quantum
  Gravity, 26, 094027

\bibitem[{{Arzoumanian} {et~al.}(2020){Arzoumanian}, {Baker}, {Blumer},
  {B{\'e}csy}, {Brazier}, {Brook}, {Burke-Spolaor}, {Chatterjee}, {Chen},
  {Cordes}, {Cornish}, {Crawford}, {Cromartie}, {Decesar}, {Demorest}, {Dolch},
  {Ellis}, {Ferrara}, {Fiore}, {Fonseca}, {Garver-Daniels}, {Gentile}, {Good},
  {Hazboun}, {Holgado}, {Islo}, {Jennings}, {Jones}, {Kaiser}, {Kaplan},
  {Kelley}, {Key}, {Laal}, {Lam}, {Lazio}, {Lorimer}, {Luo}, {Lynch},
  {Madison}, {McLaughlin}, {Mingarelli}, {Ng}, {Nice}, {Pennucci}, {Pol},
  {Ransom}, {Ray}, {Shapiro-Albert}, {Siemens}, {Simon}, {Spiewak}, {Stairs},
  {Stinebring}, {Stovall}, {Sun}, {Swiggum}, {Taylor}, {Turner}, {Vallisneri},
  {Vigeland}, {Witt}, \& {Nanograv Collaboration}}]{Arzoumanian20}
{Arzoumanian}, Z., {Baker}, P.~T., {Blumer}, H., {et~al.} 2020, \apjl, 905, L34

\bibitem[{{Arzoumanian} {et~al.}(2018){Arzoumanian}, {Baker}, {Brazier},
  {Burke-Spolaor}, {Chamberlin}, {Chatterjee}, {Christy}, {Cordes}, {Cornish},
  {Crawford}, {Thankful Cromartie}, {Crowter}, {DeCesar}, {Demorest}, {Dolch},
  {Ellis}, {Ferdman}, {Ferrara}, {Folkner}, {Fonseca}, {Garver-Daniels},
  {Gentile}, {Haas}, {Hazboun}, {Huerta}, {Islo}, {Jones}, {Jones}, {Kaplan},
  {Kaspi}, {Lam}, {Lazio}, {Levin}, {Lommen}, {Lorimer}, {Luo}, {Lynch},
  {Madison}, {McLaughlin}, {McWilliams}, {Mingarelli}, {Ng}, {Nice}, {Park},
  {Pennucci}, {Pol}, {Ransom}, {Ray}, {Rasskazov}, {Siemens}, {Simon},
  {Spiewak}, {Stairs}, {Stinebring}, {Stovall}, {Swiggum}, {Taylor},
  {Vallisneri}, {van Haasteren}, {Vigeland}, {Zhu}, \& {NANOGrav
  Collaboration}}]{Arzoumanian18}
{Arzoumanian}, Z., {Baker}, P.~T., {Brazier}, A., {et~al.} 2018, \apj, 859, 47

\bibitem[{{Ba{\~n}ados} {et~al.}(2018){Ba{\~n}ados}, {Venemans},
  {Mazzucchelli}, {Farina}, {Walter}, {Wang}, {Decarli}, {Stern}, {Fan},
  {Davies}, {Hennawi}, {Simcoe}, {Turner}, {Rix}, {Yang}, {Kelson}, {Rudie}, \&
  {Winters}}]{Banados18}
{Ba{\~n}ados}, E., {Venemans}, B.~P., {Mazzucchelli}, C., {et~al.} 2018, \nat,
  553, 473

\bibitem[{{Babak} {et~al.}(2016){Babak}, {Petiteau}, {Sesana}, {Brem},
  {Rosado}, {Taylor}, {Lassus}, {Hessels}, {Bassa}, {Burgay}, {Caballero},
  {Champion}, {Cognard}, {Desvignes}, {Gair}, {Guillemot}, {Janssen},
  {Karuppusamy}, {Kramer}, {Lazarus}, {Lee}, {Lentati}, {Liu}, {Mingarelli},
  {Os{\l}owski}, {Perrodin}, {Possenti}, {Purver}, {Sanidas}, {Smits},
  {Stappers}, {Theureau}, {Tiburzi}, {van Haasteren}, {Vecchio}, \&
  {Verbiest}}]{Babak16}
{Babak}, S., {Petiteau}, A., {Sesana}, A., {et~al.} 2016, \mnras, 455, 1665

\bibitem[{{Baldassare} {et~al.}(2020){Baldassare}, {Geha}, \&
  {Greene}}]{Baldassare20}
{Baldassare}, V.~F., {Geha}, M., \& {Greene}, J. 2020, \apj, 896, 10

\bibitem[{{Barausse} {et~al.}(2020){Barausse}, {Dvorkin}, {Tremmel},
  {Volonteri}, \& {Bonetti}}]{Barausse20}
{Barausse}, E., {Dvorkin}, I., {Tremmel}, M., {Volonteri}, M., \& {Bonetti}, M.
  2020, \apj, 904, 16

\bibitem[{{Barausse} {et~al.}(2017){Barausse}, {Shankar}, {Bernardi}, {Dubois},
  \& {Sheth}}]{Barausse17}
{Barausse}, E., {Shankar}, F., {Bernardi}, M., {Dubois}, Y., \& {Sheth}, R.~K.
  2017, \mnras, 468, 4782

\bibitem[{{Behroozi} {et~al.}(2020){Behroozi}, {Conroy}, {Wechsler}, {Hearin},
  {Williams}, {Moster}, {Yung}, {Somerville}, {Gottl{\"o}ber}, {Yepes}, \&
  {Endsley}}]{Behroozi20}
{Behroozi}, P., {Conroy}, C., {Wechsler}, R.~H., {et~al.} 2020, \mnras, 499,
  5702

\bibitem[{{Beifiori} {et~al.}(2012){Beifiori}, {Courteau}, {Corsini}, \&
  {Zhu}}]{Beifiori12}
{Beifiori}, A., {Courteau}, S., {Corsini}, E.~M., \& {Zhu}, Y. 2012, \mnras,
  419, 2497

\bibitem[{{Benson}(2010)}]{Benson10}
{Benson}, A.~J. 2010, \physrep, 495, 33

\bibitem[{{Blitz} \& {Rosolowsky}(2006)}]{Blitz06}
{Blitz}, L. \& {Rosolowsky}, E. 2006, \apj, 650, 933

\bibitem[{{Bonetti} {et~al.}(2019){Bonetti}, {Sesana}, {Haardt}, {Barausse}, \&
  {Colpi}}]{Bonetti19}
{Bonetti}, M., {Sesana}, A., {Haardt}, F., {Barausse}, E., \& {Colpi}, M. 2019,
  \mnras, 486, 4044

\bibitem[{{Bower} {et~al.}(2006){Bower}, {Benson}, {Malbon}, {Helly}, {Frenk},
  {Baugh}, {Cole}, \& {Lacey}}]{Bower06}
{Bower}, R.~G., {Benson}, A.~J., {Malbon}, R., {et~al.} 2006, \mnras, 370, 645

\bibitem[{{Cao}(2010)}]{Cao10}
{Cao}, X. 2010, \apj, 725, 388

\bibitem[{{Chen} {et~al.}(2021){Chen}, {Caballero}, {Guo}, {Chalumeau}, {Liu},
  {Shaifullah}, {Lee}, {Babak}, {Desvignes}, {Parthasarathy}, {Hu}, {van der
  Wateren}, {Antoniadis}, {Bak Nielsen}, {Bassa}, {Berthereau}, {Burgay},
  {Champion}, {Cognard}, {Falxa}, {Ferdman}, {Freire}, {Gair}, {Graikou},
  {Guillemot}, {Jang}, {Janssen}, {Karuppusamy}, {Keith}, {Kramer}, {Liu},
  {Lyne}, {Main}, {McKee}, {Mickaliger}, {Perera}, {Perrodin}, {Petiteau},
  {Porayko}, {Possenti}, {Samajdar}, {Sanidas}, {Sesana}, {Speri}, {Stappers},
  {Theureau}, {Tiburzi}, {Vecchio}, {Verbiest}, {Wang}, {Wang}, \&
  {Xu}}]{Chen21}
{Chen}, S., {Caballero}, R.~N., {Guo}, Y.~J., {et~al.} 2021, \mnras, 508, 4970

\bibitem[{{Crain} {et~al.}(2015){Crain}, {Schaye}, {Bower}, {Furlong},
  {Schaller}, {Theuns}, {Dalla Vecchia}, {Frenk}, {McCarthy}, {Helly},
  {Jenkins}, {Rosas-Guevara}, {White}, \& {Trayford}}]{Crain15}
{Crain}, R.~A., {Schaye}, J., {Bower}, R.~G., {et~al.} 2015, \mnras, 450, 1937

\bibitem[{{Croton} {et~al.}(2006){Croton}, {Springel}, {White}, {De Lucia},
  {Frenk}, {Gao}, {Jenkins}, {Kauffmann}, {Navarro}, \& {Yoshida}}]{Croton06}
{Croton}, D.~J., {Springel}, V., {White}, S. D.~M., {et~al.} 2006, \mnras, 365,
  11

\bibitem[{{Croton} {et~al.}(2016){Croton}, {Stevens}, {Tonini}, {Garel},
  {Bernyk}, {Bibiano}, {Hodkinson}, {Mutch}, {Poole}, \& {Shattow}}]{Croton16}
{Croton}, D.~J., {Stevens}, A. R.~H., {Tonini}, C., {et~al.} 2016, \apjs, 222,
  22

\bibitem[{{Dayal} {et~al.}(2019){Dayal}, {Rossi}, {Shiralilou}, {Piana},
  {Choudhury}, \& {Volonteri}}]{Dayal19}
{Dayal}, P., {Rossi}, E.~M., {Shiralilou}, B., {et~al.} 2019, \mnras, 486, 2336

\bibitem[{{De Rosa} {et~al.}(2019){De Rosa}, {Vignali}, {Bogdanovi{\'c}},
  {Capelo}, {Charisi}, {Dotti}, {Husemann}, {Lusso}, {Mayer}, {Paragi},
  {Runnoe}, {Sesana}, {Steinborn}, {Bianchi}, {Colpi}, {del Valle}, {Frey},
  {Gab{\'a}nyi}, {Giustini}, {Guainazzi}, {Haiman}, {Herrera Ruiz},
  {Herrero-Illana}, {Iwasawa}, {Komossa}, {Lena}, {Loiseau}, {Perez-Torres},
  {Piconcelli}, \& {Volonteri}}]{DeRosa19}
{De Rosa}, A., {Vignali}, C., {Bogdanovi{\'c}}, T., {et~al.} 2019, \nar, 86,
  101525

\bibitem[{{Desvignes} {et~al.}(2016){Desvignes}, {Caballero}, {Lentati},
  {Verbiest}, {Champion}, {Stappers}, {Janssen}, {Lazarus}, {Os{\l}owski},
  {Babak}, {Bassa}, {Brem}, {Burgay}, {Cognard}, {Gair}, {Graikou},
  {Guillemot}, {Hessels}, {Jessner}, {Jordan}, {Karuppusamy}, {Kramer},
  {Lassus}, {Lazaridis}, {Lee}, {Liu}, {Lyne}, {McKee}, {Mingarelli},
  {Perrodin}, {Petiteau}, {Possenti}, {Purver}, {Rosado}, {Sanidas}, {Sesana},
  {Shaifullah}, {Smits}, {Taylor}, {Theureau}, {Tiburzi}, {van Haasteren}, \&
  {Vecchio}}]{EPTA}
{Desvignes}, G., {Caballero}, R.~N., {Lentati}, L., {et~al.} 2016, \mnras, 458,
  3341

\bibitem[{{Elahi} {et~al.}(2018){Elahi}, {Welker}, {Power}, {Lagos},
  {Robotham}, {Ca{\~n}as}, \& {Poulton}}]{Elahi18}
{Elahi}, P.~J., {Welker}, C., {Power}, C., {et~al.} 2018, \mnras, 475, 5338

\bibitem[{{Ferrarese} \& {Ford}(2005)}]{Farrarese05}
{Ferrarese}, L. \& {Ford}, H. 2005, \ssr, 116, 523

\bibitem[{{Goncharov} {et~al.}(2021){Goncharov}, {Shannon}, {Reardon}, {Hobbs},
  {Zic}, {Bailes}, {Curylo}, {Dai}, {Kerr}, {Lower}, {Manchester}, {Mandow},
  {Middleton}, {Miles}, {Parthasarathy}, {Thrane}, {Thyagarajan}, {Xue}, {Zhu},
  {Cameron}, {Feng}, {Luo}, {Russell}, {Sarkissian}, {Spiewak}, {Wang}, {Wang},
  {Zhang}, \& {Zhang}}]{Goncharov21}
{Goncharov}, B., {Shannon}, R.~M., {Reardon}, D.~J., {et~al.} 2021, arXiv
  e-prints, arXiv:2107.12112

\bibitem[{{G{\"u}ltekin} {et~al.}(2009){G{\"u}ltekin}, {Richstone}, {Gebhardt},
  {Lauer}, {Tremaine}, {Aller}, {Bender}, {Dressler}, {Faber}, {Filippenko},
  {Green}, {Ho}, {Kormendy}, {Magorrian}, {Pinkney}, \& {Siopis}}]{Gultekin09}
{G{\"u}ltekin}, K., {Richstone}, D.~O., {Gebhardt}, K., {et~al.} 2009, \apj,
  698, 198

\bibitem[{{Habouzit} {et~al.}(2021){Habouzit}, {Li}, {Somerville}, {Genel},
  {Pillepich}, {Volonteri}, {Dav{\'e}}, {Rosas-Guevara}, {McAlpine}, {Peirani},
  {Hernquist}, {Angl{\'e}s-Alc{\'a}zar}, {Reines}, {Bower}, {Dubois}, {Nelson},
  {Pichon}, \& {Vogelsberger}}]{Habouzit20}
{Habouzit}, M., {Li}, Y., {Somerville}, R.~S., {et~al.} 2021, \mnras, 503, 1940

\bibitem[{{Henriques} {et~al.}(2013){Henriques}, {White}, {Thomas}, {Angulo},
  {Guo}, {Lemson}, \& {Springel}}]{Henriques13}
{Henriques}, B. M.~B., {White}, S. D.~M., {Thomas}, P.~A., {et~al.} 2013,
  \mnras, 431, 3373

\bibitem[{{Husa} {et~al.}(2016){Husa}, {Khan}, {Hannam}, {P{\"u}rrer}, {Ohme},
  {Forteza}, \& {Boh{\'e}}}]{Husa16}
{Husa}, S., {Khan}, S., {Hannam}, M., {et~al.} 2016, \prd, 93, 044006

\bibitem[{{Inayoshi} {et~al.}(2020){Inayoshi}, {Visbal}, \&
  {Haiman}}]{Inayoshi20}
{Inayoshi}, K., {Visbal}, E., \& {Haiman}, Z. 2020, \araa, 58, 27

\bibitem[{{Izquierdo-Villalba} {et~al.}(2022){Izquierdo-Villalba}, {Sesana},
  {Bonoli}, \& {Colpi}}]{Villalba22}
{Izquierdo-Villalba}, D., {Sesana}, A., {Bonoli}, S., \& {Colpi}, M. 2022,
  \mnras, 509, 3488

\bibitem[{{Joshi} {et~al.}(2018){Joshi}, {Arumugasamy}, {Bagchi},
  {Bandyopadhyay}, {Basu}, {Dhanda Batra}, {Bethapudi}, {Choudhary}, {De},
  {Dey}, {Gopakumar}, {Gupta}, {Krishnakumar}, {Maan}, {Manoharan}, {Naidu},
  {Nandi}, {Pathak}, {Surnis}, \& {Susobhanan}}]{INPTA}
{Joshi}, B.~C., {Arumugasamy}, P., {Bagchi}, M., {et~al.} 2018, Journal of
  Astrophysics and Astronomy, 39, 51

\bibitem[{{Katz} {et~al.}(2020){Katz}, {Kelley}, {Dosopoulou}, {Berry},
  {Blecha}, \& {Larson}}]{Katz20}
{Katz}, M.~L., {Kelley}, L.~Z., {Dosopoulou}, F., {et~al.} 2020, \mnras, 491,
  2301

\bibitem[{{Katz} \& {Larson}(2019)}]{Katz19}
{Katz}, M.~L. \& {Larson}, S.~L. 2019, \mnras, 483, 3108

\bibitem[{{Kauffmann} \& {Haehnelt}(2000)}]{Kauffman00}
{Kauffmann}, G. \& {Haehnelt}, M. 2000, \mnras, 311, 576

\bibitem[{{Kelley} {et~al.}(2017{\natexlab{a}}){Kelley}, {Blecha}, \&
  {Hernquist}}]{Kelley17a}
{Kelley}, L.~Z., {Blecha}, L., \& {Hernquist}, L. 2017{\natexlab{a}}, \mnras,
  464, 3131

\bibitem[{{Kelley} {et~al.}(2017{\natexlab{b}}){Kelley}, {Blecha}, {Hernquist},
  {Sesana}, \& {Taylor}}]{Kelley17b}
{Kelley}, L.~Z., {Blecha}, L., {Hernquist}, L., {Sesana}, A., \& {Taylor},
  S.~R. 2017{\natexlab{b}}, \mnras, 471, 4508

\bibitem[{{Kelly} \& {Merloni}(2012)}]{Kelly12}
{Kelly}, B.~C. \& {Merloni}, A. 2012, Advances in Astronomy, 2012, 970858

\bibitem[{{Khan} {et~al.}(2016){Khan}, {Husa}, {Hannam}, {Ohme}, {P{\"u}rrer},
  {Forteza}, \& {Boh{\'e}}}]{Khan16}
{Khan}, S., {Husa}, S., {Hannam}, M., {et~al.} 2016, \prd, 93, 044007

\bibitem[{{Kulier} {et~al.}(2015){Kulier}, {Ostriker}, {Natarajan}, {Lackner},
  \& {Cen}}]{Kulier15}
{Kulier}, A., {Ostriker}, J.~P., {Natarajan}, P., {Lackner}, C.~N., \& {Cen},
  R. 2015, \apj, 799, 178

\bibitem[{{Lacey} \& {Cole}(1993)}]{Lacey93}
{Lacey}, C. \& {Cole}, S. 1993, \mnras, 262, 627

\bibitem[{{Lacey} {et~al.}(2016){Lacey}, {Baugh}, {Frenk}, {Benson}, {Bower},
  {Cole}, {Gonzalez-Perez}, {Helly}, {Lagos}, \& {Mitchell}}]{Lacey16}
{Lacey}, C.~G., {Baugh}, C.~M., {Frenk}, C.~S., {et~al.} 2016, \mnras, 462,
  3854

\bibitem[{{Lagos} {et~al.}(2013){Lagos}, {Lacey}, \& {Baugh}}]{Lagos13}
{Lagos}, C. d.~P., {Lacey}, C.~G., \& {Baugh}, C.~M. 2013, \mnras, 436, 1787

\bibitem[{{Lagos} {et~al.}(2018){Lagos}, {Tobar}, {Robotham}, {Obreschkow},
  {Mitchell}, {Power}, \& {Elahi}}]{Lagos18}
{Lagos}, C. d.~P., {Tobar}, R.~J., {Robotham}, A. S.~G., {et~al.} 2018, \mnras,
  481, 3573

\bibitem[{{L{\"a}sker} {et~al.}(2014){L{\"a}sker}, {Ferrarese}, {van de Ven},
  \& {Shankar}}]{Lasker14}
{L{\"a}sker}, R., {Ferrarese}, L., {van de Ven}, G., \& {Shankar}, F. 2014,
  \apj, 780, 70

\bibitem[{{Lee}(2016)}]{CPTA}
{Lee}, K.~J. 2016, in Astronomical Society of the Pacific Conference Series,
  Vol. 502, Frontiers in Radio Astronomy and FAST Early Sciences Symposium
  2015, ed. L.~{Qain} \& D.~{Li}, 19

\bibitem[{{Leroy} {et~al.}(2013){Leroy}, {Walter}, {Sandstrom}, {Schruba},
  {Munoz-Mateos}, {Bigiel}, {Bolatto}, {Brinks}, {de Blok}, {Meidt}, {Rix},
  {Rosolowsky}, {Schinnerer}, {Schuster}, \& {Usero}}]{Leroy13}
{Leroy}, A.~K., {Walter}, F., {Sandstrom}, K., {et~al.} 2013, \aj, 146, 19

\bibitem[{{Maiolino} {et~al.}(2012){Maiolino}, {Gallerani}, {Neri}, {Cicone},
  {Ferrara}, {Genzel}, {Lutz}, {Sturm}, {Tacconi}, {Walter}, {Feruglio},
  {Fiore}, \& {Piconcelli}}]{Maiolino12}
{Maiolino}, R., {Gallerani}, S., {Neri}, R., {et~al.} 2012, \mnras, 425, L66

\bibitem[{{Manchester} {et~al.}(2013){Manchester}, {Hobbs}, {Bailes}, {Coles},
  {van Straten}, {Keith}, {Shannon}, {Bhat}, {Brown}, {Burke-Spolaor},
  {Champion}, {Chaudhary}, {Edwards}, {Hampson}, {Hotan}, {Jameson}, {Jenet},
  {Kesteven}, {Khoo}, {Kocz}, {Maciesiak}, {Oslowski}, {Ravi}, {Reynolds},
  {Sarkissian}, {Verbiest}, {Wen}, {Wilson}, {Yardley}, {Yan}, \& {You}}]{PPTA}
{Manchester}, R.~N., {Hobbs}, G., {Bailes}, M., {et~al.} 2013, \pasa, 30, e017

\bibitem[{{Mazzucchelli} {et~al.}(2017){Mazzucchelli}, {Ba{\~n}ados},
  {Venemans}, {Decarli}, {Farina}, {Walter}, {Eilers}, {Rix}, {Simcoe},
  {Stern}, {Fan}, {Schlafly}, {De Rosa}, {Hennawi}, {Chambers}, {Greiner},
  {Burgett}, {Draper}, {Kaiser}, {Kudritzki}, {Magnier}, {Metcalfe}, {Waters},
  \& {Wainscoat}}]{Mazzucchelli17}
{Mazzucchelli}, C., {Ba{\~n}ados}, E., {Venemans}, B.~P., {et~al.} 2017, \apj,
  849, 91

\bibitem[{{McConnell} \& {Ma}(2013)}]{McConnell13}
{McConnell}, N.~J. \& {Ma}, C.-P. 2013, \apj, 764, 184

\bibitem[{{McLaughlin}(2013)}]{NANOGRAV}
{McLaughlin}, M.~A. 2013, Classical and Quantum Gravity, 30, 224008

\bibitem[{{Merloni} \& {Heinz}(2008)}]{Merloni08}
{Merloni}, A. \& {Heinz}, S. 2008, \mnras, 388, 1011

\bibitem[{{Middleton} {et~al.}(2021){Middleton}, {Sesana}, {Chen}, {Vecchio},
  {Del Pozzo}, \& {Rosado}}]{Middleton21}
{Middleton}, H., {Sesana}, A., {Chen}, S., {et~al.} 2021, \mnras, 502, L99

\bibitem[{{Nandra} {et~al.}(2013){Nandra}, {Barret}, {Barcons}, {Fabian}, {den
  Herder}, {Piro}, {Watson}, {Adami}, {Aird}, {Afonso}, \& et~al.}]{Athena13}
{Nandra}, K., {Barret}, D., {Barcons}, X., {et~al.} 2013, arXiv e-prints,
  arXiv:1306.2307

\bibitem[{{Nguyen} {et~al.}(2019){Nguyen}, {Seth}, {Neumayer}, {Iguchi},
  {Cappellari}, {Strader}, {Chomiuk}, {Tremou}, {Pacucci}, {Nakanishi},
  {Bahramian}, {Nguyen}, {den Brok}, {Ahn}, {Voggel}, {Kacharov}, {Tsukui},
  {Ly}, {Dumont}, \& {Pechetti}}]{Nguyen19}
{Nguyen}, D.~D., {Seth}, A.~C., {Neumayer}, N., {et~al.} 2019, \apj, 872, 104

\bibitem[{{Nulsen} \& {Fabian}(2000)}]{Nulsen00}
{Nulsen}, P.~E.~J. \& {Fabian}, A.~C. 2000, \mnras, 311, 346

\bibitem[{{Parkinson} {et~al.}(2008){Parkinson}, {Cole}, \&
  {Helly}}]{Parkinson08}
{Parkinson}, H., {Cole}, S., \& {Helly}, J. 2008, \mnras, 383, 557

\bibitem[{{Planck Collaboration} {et~al.}(2016){Planck Collaboration}, {Ade},
  {Aghanim}, {Arnaud}, {Ashdown}, {Aumont}, {Baccigalupi}, {Banday},
  {Barreiro}, {Bartlett}, {Bartolo}, {Battaner}, {Battye}, {Benabed},
  {Beno{\^\i}t}, {Benoit-L{\'e}vy}, {Bernard}, {Bersanelli}, {Bielewicz},
  {Bock}, {Bonaldi}, {Bonavera}, {Bond}, {Borrill}, {Bouchet}, {Boulanger},
  {Bucher}, {Burigana}, {Butler}, {Calabrese}, {Cardoso}, {Catalano},
  {Challinor}, {Chamballu}, {Chary}, {Chiang}, {Chluba}, {Christensen},
  {Church}, {Clements}, {Colombi}, {Colombo}, {Combet}, {Coulais}, {Crill},
  {Curto}, {Cuttaia}, {Danese}, {Davies}, {Davis}, {de Bernardis}, {de Rosa},
  {de Zotti}, {Delabrouille}, {D{\'e}sert}, {Di Valentino}, {Dickinson},
  {Diego}, {Dolag}, {Dole}, {Donzelli}, {Dor{\'e}}, {Douspis}, {Ducout},
  {Dunkley}, {Dupac}, {Efstathiou}, {Elsner}, {En{\ss}lin}, {Eriksen},
  {Farhang}, {Fergusson}, {Finelli}, {Forni}, {Frailis}, {Fraisse},
  {Franceschi}, {Frejsel}, {Galeotta}, {Galli}, {Ganga}, {Gauthier}, {Gerbino},
  {Ghosh}, {Giard}, {Giraud-H{\'e}raud}, {Giusarma}, {Gjerl{\o}w},
  {Gonz{\'a}lez-Nuevo}, {G{\'o}rski}, {Gratton}, {Gregorio}, {Gruppuso},
  {Gudmundsson}, {Hamann}, {Hansen}, {Hanson}, {Harrison}, {Helou},
  {Henrot-Versill{\'e}}, {Hern{\'a}ndez-Monteagudo}, {Herranz}, {Hildebrandt},
  {Hivon}, {Hobson}, {Holmes}, {Hornstrup}, {Hovest}, {Huang}, {Huffenberger},
  {Hurier}, {Jaffe}, {Jaffe}, {Jones}, {Juvela}, {Keih{\"a}nen}, {Keskitalo},
  {Kisner}, {Kneissl}, {Knoche}, {Knox}, {Kunz}, {Kurki-Suonio}, {Lagache},
  {L{\"a}hteenm{\"a}ki}, {Lamarre}, {Lasenby}, {Lattanzi}, {Lawrence}, {Leahy},
  {Leonardi}, {Lesgourgues}, {Levrier}, {Lewis}, {Liguori}, {Lilje},
  {Linden-V{\o}rnle}, {L{\'o}pez-Caniego}, {Lubin}, {Mac{\'\i}as-P{\'e}rez},
  {Maggio}, {Maino}, {Mandolesi}, {Mangilli}, {Marchini}, {Maris}, {Martin},
  {Martinelli}, {Mart{\'\i}nez-Gonz{\'a}lez}, {Masi}, {Matarrese}, {McGehee},
  {Meinhold}, {Melchiorri}, {Melin}, {Mendes}, {Mennella}, {Migliaccio},
  {Millea}, {Mitra}, {Miville-Desch{\^e}nes}, {Moneti}, {Montier}, {Morgante},
  {Mortlock}, {Moss}, {Munshi}, {Murphy}, {Naselsky}, {Nati}, {Natoli},
  {Netterfield}, {N{\o}rgaard-Nielsen}, {Noviello}, {Novikov}, {Novikov},
  {Oxborrow}, {Paci}, {Pagano}, {Pajot}, {Paladini}, {Paoletti}, {Partridge},
  {Pasian}, {Patanchon}, {Pearson}, {Perdereau}, {Perotto}, {Perrotta},
  {Pettorino}, {Piacentini}, {Piat}, {Pierpaoli}, {Pietrobon}, {Plaszczynski},
  {Pointecouteau}, {Polenta}, {Popa}, {Pratt}, {Pr{\'e}zeau}, {Prunet},
  {Puget}, {Rachen}, {Reach}, {Rebolo}, {Reinecke}, {Remazeilles}, {Renault},
  {Renzi}, {Ristorcelli}, {Rocha}, {Rosset}, {Rossetti}, {Roudier},
  {Rouill{\'e} d'Orfeuil}, {Rowan-Robinson}, {Rubi{\~n}o-Mart{\'\i}n},
  {Rusholme}, {Said}, {Salvatelli}, {Salvati}, {Sandri}, {Santos},
  {Savelainen}, {Savini}, {Scott}, {Seiffert}, {Serra}, {Shellard}, {Spencer},
  {Spinelli}, {Stolyarov}, {Stompor}, {Sudiwala}, {Sunyaev}, {Sutton},
  {Suur-Uski}, {Sygnet}, {Tauber}, {Terenzi}, {Toffolatti}, {Tomasi},
  {Tristram}, {Trombetti}, {Tucci}, {Tuovinen}, {T{\"u}rler}, {Umana},
  {Valenziano}, {Valiviita}, {Van Tent}, {Vielva}, {Villa}, {Wade}, {Wandelt},
  {Wehus}, {White}, {White}, {Wilkinson}, {Yvon}, {Zacchei}, \&
  {Zonca}}]{Planck}
{Planck Collaboration}, {Ade}, P.~A.~R., {Aghanim}, N., {et~al.} 2016, \aap,
  594, A13

\bibitem[{{Press} \& {Schechter}(1974)}]{Press74}
{Press}, W.~H. \& {Schechter}, P. 1974, \apj, 187, 425

\bibitem[{{Reines} {et~al.}(2013){Reines}, {Greene}, \& {Geha}}]{Reines13}
{Reines}, A.~E., {Greene}, J.~E., \& {Geha}, M. 2013, \apj, 775, 116

\bibitem[{{Reines} \& {Volonteri}(2015)}]{Reines15}
{Reines}, A.~E. \& {Volonteri}, M. 2015, \apj, 813, 82

\bibitem[{{Robson} {et~al.}(2019){Robson}, {Cornish}, \& {Liu}}]{Robson19}
{Robson}, T., {Cornish}, N.~J., \& {Liu}, C. 2019, Classical and Quantum
  Gravity, 36, 105011

\bibitem[{{Rosado} {et~al.}(2015){Rosado}, {Sesana}, \& {Gair}}]{Rosado15}
{Rosado}, P.~A., {Sesana}, A., \& {Gair}, J. 2015, \mnras, 451, 2417

\bibitem[{{Saade} {et~al.}(2020){Saade}, {Stern}, {Brightman}, {Haiman},
  {Djorgovski}, {D'Orazio}, {Ford}, {Graham}, {Jun}, {Kraft}, {McKernan},
  {Vikhlinin}, \& {Walton}}]{Saade20}
{Saade}, M.~L., {Stern}, D., {Brightman}, M., {et~al.} 2020, \apj, 900, 148

\bibitem[{{Salcido} {et~al.}(2016){Salcido}, {Bower}, {Theuns}, {McAlpine},
  {Schaller}, {Crain}, {Schaye}, \& {Regan}}]{Salcido16}
{Salcido}, J., {Bower}, R.~G., {Theuns}, T., {et~al.} 2016, \mnras, 463, 870

\bibitem[{{Schaye} {et~al.}(2015){Schaye}, {Crain}, {Bower}, {Furlong},
  {Schaller}, {Theuns}, {Dalla Vecchia}, {Frenk}, {McCarthy}, {Helly},
  {Jenkins}, {Rosas-Guevara}, {White}, {Baes}, {Booth}, {Camps}, {Navarro},
  {Qu}, {Rahmati}, {Sawala}, {Thomas}, \& {Trayford}}]{Schaye15}
{Schaye}, J., {Crain}, R.~A., {Bower}, R.~G., {et~al.} 2015, \mnras, 446, 521

\bibitem[{{Sesana} {et~al.}(2011){Sesana}, {Gair}, {Berti}, \&
  {Volonteri}}]{Sesana11}
{Sesana}, A., {Gair}, J., {Berti}, E., \& {Volonteri}, M. 2011, \prd, 83,
  044036

\bibitem[{{Shankar} {et~al.}(2019){Shankar}, {Bernardi}, {Richardson},
  {Marsden}, {Sheth}, {Allevato}, {Graziani}, {Mezcua}, {Ricci}, {Penny}, {La
  Franca}, \& {Pacucci}}]{Shankar19}
{Shankar}, F., {Bernardi}, M., {Richardson}, K., {et~al.} 2019, \mnras, 485,
  1278

\bibitem[{{Shankar} {et~al.}(2009){Shankar}, {Weinberg}, \&
  {Miralda-Escud{\'e}}}]{Shankar09}
{Shankar}, F., {Weinberg}, D.~H., \& {Miralda-Escud{\'e}}, J. 2009, \apj, 690,
  20

\bibitem[{{Shannon} {et~al.}(2015){Shannon}, {Ravi}, {Lentati}, {Lasky},
  {Hobbs}, {Kerr}, {Manchester}, {Coles}, {Levin}, {Bailes}, {Bhat},
  {Burke-Spolaor}, {Dai}, {Keith}, {Os{\l}owski}, {Reardon}, {van Straten},
  {Toomey}, {Wang}, {Wen}, {Wyithe}, \& {Zhu}}]{Shannon15}
{Shannon}, R.~M., {Ravi}, V., {Lentati}, L.~T., {et~al.} 2015, Science, 349,
  1522

\bibitem[{{Sobacchi} \& {Mesinger}(2013)}]{Sobacchi13}
{Sobacchi}, E. \& {Mesinger}, A. 2013, \mnras, 432, L51

\bibitem[{{Taylor}(2000)}]{Taylor00}
{Taylor}, A.~R. 2000, in Perspectives on Radio Astronomy: Science with Large
  Antenna Arrays, ed. M.~P. {van Haarlem}, 1

\bibitem[{{Vogelsberger} {et~al.}(2014){Vogelsberger}, {Genel}, {Springel},
  {Torrey}, {Sijacki}, {Xu}, {Snyder}, {Nelson}, \& {Hernquist}}]{Illustris}
{Vogelsberger}, M., {Genel}, S., {Springel}, V., {et~al.} 2014, \mnras, 444,
  1518

\bibitem[{{Volonteri}(2010)}]{Volonteri10}
{Volonteri}, M. 2010, \aapr, 18, 279

\bibitem[{{Volonteri} {et~al.}(2021){Volonteri}, {Habouzit}, \&
  {Colpi}}]{Volonteri21}
{Volonteri}, M., {Habouzit}, M., \& {Colpi}, M. 2021, Nature Reviews Physics,
  3, 732

\bibitem[{{Wu} {et~al.}(2015){Wu}, {Wang}, {Fan}, {Yi}, {Zuo}, {Bian}, {Jiang},
  {McGreer}, {Wang}, {Yang}, {Yang}, {Thompson}, \& {Beletsky}}]{Wu15}
{Wu}, X.-B., {Wang}, F., {Fan}, X., {et~al.} 2015, \nat, 518, 512

\bibitem[{{Yue} {et~al.}(2014){Yue}, {Ferrara}, {Salvaterra}, {Xu}, \&
  {Chen}}]{Yue14}
{Yue}, B., {Ferrara}, A., {Salvaterra}, R., {Xu}, Y., \& {Chen}, X. 2014,
  \mnras, 440, 1263

\end{thebibliography}

\end{document}